\documentclass[useAMS,usenatbib]{mn2e}

\usepackage{aecompl}
\usepackage[toc,page]{appendix}

\usepackage{graphicx}
\usepackage{amssymb}
\usepackage{amsmath}
\usepackage{aas_macros}
\usepackage{deluxetable}
\usepackage{lscape}
\usepackage{rotating}
\usepackage[symbol*]{footmisc}
\usepackage{gensymb}
\usepackage{tikz}



\title[The young, cold and quiescent GMF G214.5-1.8]{A Herschel study of G214.5-1.8: a young, cold and quiescent giant molecular filament on the shell of a HI superbubble}

\author[S. D. Clarke et al.]{S. D. Clarke$^{1}$\thanks{E-mail: sclarke@asiaa.sinica.edu.tw }, {\'A} S{\'a}nchez-Monge$^{2}$, G. M. Williams$^{3}$, A. D. P. Howard$^{4}$,\newauthor S. Walch$^{2,5}$ and N. Schneider$^{2}$.\\$^{1}$Academia Sinica, Institute of Astronomy and Astrophysics, Taipei, Taiwan\\$^{2}$I. Physikalisches Institut, Universit{\"a}t zu K{\"o}ln, Z{\"u}lpicher Str. 77, D-50937 K{\"o}ln, Germany \\$^{3}$School of Physics and Astronomy, University of Leeds, Leeds, LS2 9JT, UK \\$^{4}$School of Physics and Astronomy, Cardiff University, Cardiff, CF24 3AA, UK \\$^{5}$Cologne Centre for Data and Simulation Science, University of Cologne, Cologne, Germany \thanks{www.cds.uni-koeln.de}}

\begin{document}

\date{}

\pagerange{\pageref{firstpage}--\pageref{lastpage}} \pubyear{2002}

\maketitle

\label{firstpage}

\begin{abstract}
We present an analysis of the outer Galaxy giant molecular filament (GMF) G214.5-1.8 (G214.5) using Herschel data. We find that G214.5 has a mass of $\sim$ 16,000 M$_{\odot}$, yet hosts only 15 potentially protostellar 70 $\mu$m sources, making it highly quiescent compared to equally massive clouds such as Serpens and Mon R2. We show that G214.5 has a unique morphology, consisting of a narrow `Main filament' running north-south and a perpendicular `Head' structure running east-west. We identify 33 distinct massive clumps from the column density maps, 8 of which are protostellar. However, the star formation activity is not evenly spread across G214.5 but rather predominantly located in the Main filament. Studying the Main filament in a manner similar to previous works, we find that G214.5 is most like a 'Bone' candidate GMF, highly elongated and massive, but it is colder and narrower than any such GMF. It also differs significantly due to its low fraction of high column density gas. Studying the radial profile, we discover that G214.5 is highly asymmetric and resembles filaments which are known to be compressed externally. Considering its environment, we find that G214.5 is co-incident, spatially and kinematically, with a HI superbubble. We discuss how a potential interaction between G214.5 and the superbubble may explain G214.5's morphology, asymmetry and, paucity of dense gas and star formation activity, highlighting the intersection of a bubble-driven interstellar medium paradigm with that of a filament paradigm for star formation.
\end{abstract}


\section{Introduction}\label{SEC:INTRO}%

It has long been known that filaments have been connected to the star formation process and contain large quantities of high-density molecular gas \citep{Bar1907,SchElm79}. The Herschel Space Observatory sparked renewed interest in filaments due to their ubiquity in observations across numerous environments \citep{And10,Arz13,Kon15,Cox16,Mar16,Mie18,Wil18,Arz19,How19,Suri19,Bon20a,Bon20b,Zav20}. As a consequence, there has been a number of recent theoretical works studying the stability, fragmentation and collapse of filaments to better understand what role they play in star formation \citep{FisMar12,Hei13,Smi14,Gom14,Cla15,Hei16,Cla16,Cla17,Zam17,Hei18,Cla18,Li19,Smi20,Cla20}.

In the past decade a class of large-scale ($\gtrapprox 10$ pc) filaments called Giant Molecular Filaments (GMFs) has been studied \citep{Jac10,Goo14,Rag14,Wang15,Abr16,Wang16,Zuc18,Zha19,Wang20,Col21}. These objects are of interest due to the fact they are sufficiently long as to be shaped by the large-scale environment and Galactic dynamics, whilst also being active star forming sites; thus illuming the influence of the large-scale ISM on the star formation process. It is currently thought that the majority of these GMFs are formed due to a combination of shear and the Galactic potential elongating gas and aligning it with spiral arms \citep{,Dua16,Dua17,Smi20}.

\citet{Zuc18} collected data from multiple studies to produce a standardized catalogue of 45 GMFs in the inner Galaxy. Using the aspect ratio and the cold \& high column density fraction of the clouds, \citet{Zuc18} identify three categories: Elongated Giant Molecular Clouds with low aspect ratios ($\sim 8$) and low cold \& high column density fractions ($< 10\%$); Elongated Dense Core Complexes with moderate aspect ratios ($\sim 10$) and high cold \& high column density fractions ($10 - 75 \%$); and 'Bone' candidates with high aspect ratios ($>20$) and moderate cold \& high column density fraction ($10-50\%$). The \citet{Zuc18} catalogue is complemented by the recent work of \citet{Col21} which provides a catalogue of 37 GMFs in the outer Galaxy; however, they use only $^{12}$CO (2-1) data and so cannot determine the cold \& high column density fraction. \citet{Col21} find that on average GMFs in the outer Galaxy have masses and line-masses an order of magnitude smaller than their inner Galaxy counterpoints while having comparable sizes and aspect ratios. Also focused on the inner Galaxy, \citet{Zha19} collate a total of 57 GMFs using the GRS and ThrUMMS CO surveys, studying their star formation rates and efficiencies using near- and mid-infrared data. 

To further this study of GMFs, especially in the outer Galaxy, we present a comprehensive study using Herschel data of the GMF G214.5-1.8. The paper is structured in the following manner: in section \ref{SEC:G214} we describe G214.5-1.8 and previous studies of it; in section \ref{SEC:OBS} we discuss the Herschel observational data used; in section \ref{SEC:RES} we present the column density and temperature maps of G214.5-1.8, as well as a catalogue of its 70 $\mu$m sources; in section \ref{SEC:DIS} we study the properties of the clumps and the main filament of G214.5-1.8 and compare to observational and theoretical results in the literature; in section \ref{SEC:BUB} we argue that G214.5-1.8 lies on the shell of a HI superbubble which impacts its evolution and star formation; and in section \ref{SEC:CON} we summarize our results.

\section{The giant molecular filament G214.5-1.8}\label{SEC:G214}%

G214.5-1.8 (hereafter G214.5) is a satellite cloud associated with the larger Maddalena's cloud in the 3rd galactic quadrant, lying at a distance of approximately 2.3 kpc towards the outer Galaxy \citep{MadTha85,Lee91,Yan19}. The distance estimate comes from considering the extinction of GAIA identified stars and has an error of 2301$^{+142}_{-149}$pc \citep{Yan19,Zuc20}, and is consistent with the kinematic distance estimates which have larger errors, 2620$^{+840}_{-740}$pc \citep{Yan19} and 2540$\pm758$pc \citep{Mege21}. Note that G214.5 has sometimes been referred to as S287-North as this is the name given to the bright IRAS source contained within the cloud, IRAS 06453-0205. Both Maddalena's and it satellite clouds, including G214.5, have low CO excitation temperatures and are rather quiescent when compared to similarly sized inner Galaxy counterparts \citep{Lee91}.

\begin{figure*}
\centering
\includegraphics[width=0.99\linewidth]{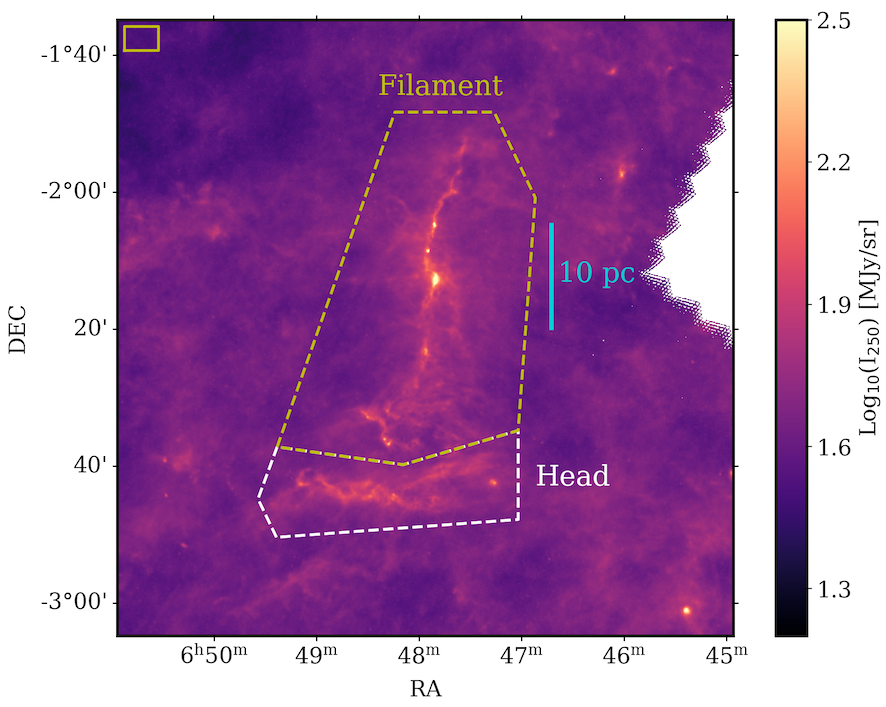}
\caption{A plot showing the 250 $\mu$m emission covering the G214.5-1.8 giant molecular filament and its environment. G214.5-1.8 is enclosed by the dashed yellow and white polygons, which is further sub-divided into the main filament (yellow) and the head structure (white). The solid yellow rectangle in the top left shows the region used to determine the contamination of the column density in section \ref{SSEC:CDTEMP}. The blue scale shows the apparent size of 10 pc at a distance of 2.3 kpc.}
\label{fig::Bound}
\end{figure*}

G214.5 can be seen in figure \ref{fig::Bound}, with Maddalena's cloud lying to the south out of the field of view. One can see that G214.5 has a unique morphology, consisting of two distinct regions: a long and thin filament seen lying north to south and a wide structure lying east to west. We refer to these two regions as the main filament and the head structure respectively. The dashed polygons in figure \ref{fig::Bound} show the boundary definitions of these structures which we use in this work. We define the combined area of these regions as the GMF G214.5-1.8 and the area outside the polygons as the cloud's environment/surroundings. The two structures, the main filament and the head structure, are velocity coherent \citep{Lee94} when observed in $^{12}$CO and $^{13}$CO, meaning that it is highly unlikely for them to be two unrelated structures along the line-of-sight. There exists a smooth velocity gradient from the southern to the northern tip, from $\sim22$ km/s to $\sim32$ km/s. The length of G214.5, from north to south, is approximately 35 pc and the width of the head structure is approximately 20 pc.

Due to its location in the outer Galaxy, the cloud G214.5 has not been covered in most CO Galactic surveys; however, G214.5 does host the bright source IRAS 06453-0209 (seen in roughly the centre of the main filament region) which has been included as a target in surveys looking for signs of star formation. These have included surveys looking for CO molecular outflows, K-band point sources, H$_2$O masers as well as class I and II methanol masers \citep{Fuk89,Fel92,Hod94,Cod95,Lee96,Lar99,Fur03,Sun07}. The results show that there exists a bipolar outflow emanating from this source which is associated with at least two stars in the K-band. There have been no detections of maser emission. This has lead to the conclusion that the star formation in the source is currently limited to only low-mass star formation despite its brightness \citep[L$_{\rm IR} \sim 230$ L$_{\odot}$,][]{Lar99}.

Recently the MWISP CO project has covered a large area of the third quadrant which includes G214.5 \citep[for details on the project see][]{Su19}. The MWISP data from the region $209.75\degree \le l \le 219.75\degree$ and $|b| \le 5\degree$ was studied by \citet{Yan19} and G214.5 was identified as a distinct cloud using a dendrogram analysis (their cloud 7). They confirm a distance of $\sim$ 2.3 kpc and the velocity coherence of the structure. Moreover, using an X$_{\rm CO}$ factor \citet{Yan19} estimate that G214.5 has a mass of $2.6 \times 10^{4}$ M$_{\odot}$. 

With its distinct morphology, high mass and apparent lack of high-mass star formation, G214.5 is an interesting object to study and provides an example of a GMF in the outer Galaxy.

\section{Observations}\label{SEC:OBS}%
The \textit{Herschel} observations of G214.5 were taken as part of the {\em Herschel} infrared Galactic Plane Survey \cite[Hi-GAL]{Mol10}. They consist of $70\mu{\rm m}$ and $160\mu{\rm m}$ data from PACS \citep{Pog10}, and $250\mu{\rm m}$, $350\mu{\rm m}$ and $500\mu{\rm m}$ data from SPIRE \citep{Gri10}. The scans were taken in the fast scan ($60''\,\rm{s}^{-1}$) PACS/SPIRE parallel mode, and two pairs of nominal and orthonormal scans, one pair observing the northern part of the structure (observation IDs 1342245150 and 1342245151), and the other pair observing the southern part (observation IDs 1342253429 and 1342253430).

The observations were reduced using the HIPE User Release v15.0.1. PACS maps were produced using a modified version of the \verb|JScanam| task, whilst SPIRE maps were produced with the \verb|mosaic| script operating on Level 2 data products. We determined the Zero-Point Offsets for the PACS maps through comparison with \textit{Planck} and \textsc{iras} observations of the same region \citep[cf.][]{Ber10}. The adopted offsets were $-5.4\,\rm{MJy\,sr^{-1}}$ and $59.4\,\rm{MJy\,sr^{-1}}$ for the {$70\mu{\rm m}$ and $160\mu{\rm m}$} bands, respectively. The SPIRE Zero-Point Offsets were applied automatically as part of the HIPE reduction process.

The {\sc fwhm}s of the circular beam profiles are 8.5'', 13.5'', 18.2'', 24.9'', and 36.3'', for -- respectively -- the $70\mu{\rm m}$, $160\mu{\rm m}$, $250\mu{\rm m}$, $350\mu{\rm m}$ and $500\mu{\rm m}$ wavebands (Herschel Explanatory Supplement Vol. III, Vol. IV). In reality, the PACS beams are distorted by the fast scan PACS/SPIRE parallel mode, producing effective non-circular beamsizes of $\sim 6'' \times 12''$ for the $70\mu{\rm m}$ waveband, and $\sim 12'' \times 16''$ for the $160\mu{\rm m}$ waveband, but this is not generally considered due to the addition of the cross scans which lead to an averaging of the beam profiles. 

The final images from each band are placed on identical grids with a size of 1.5 $\degree$ by 1.5 $\degree$, and a pixel size of 2''. These images can be seen in appendix \ref{APP:HBAND}, as well as an RGB image (R: 500 $\mu$m, G: 250 $\mu$m, B: 160 $\mu$m).

\begin{figure}
\centering
\includegraphics[width=0.95\linewidth]{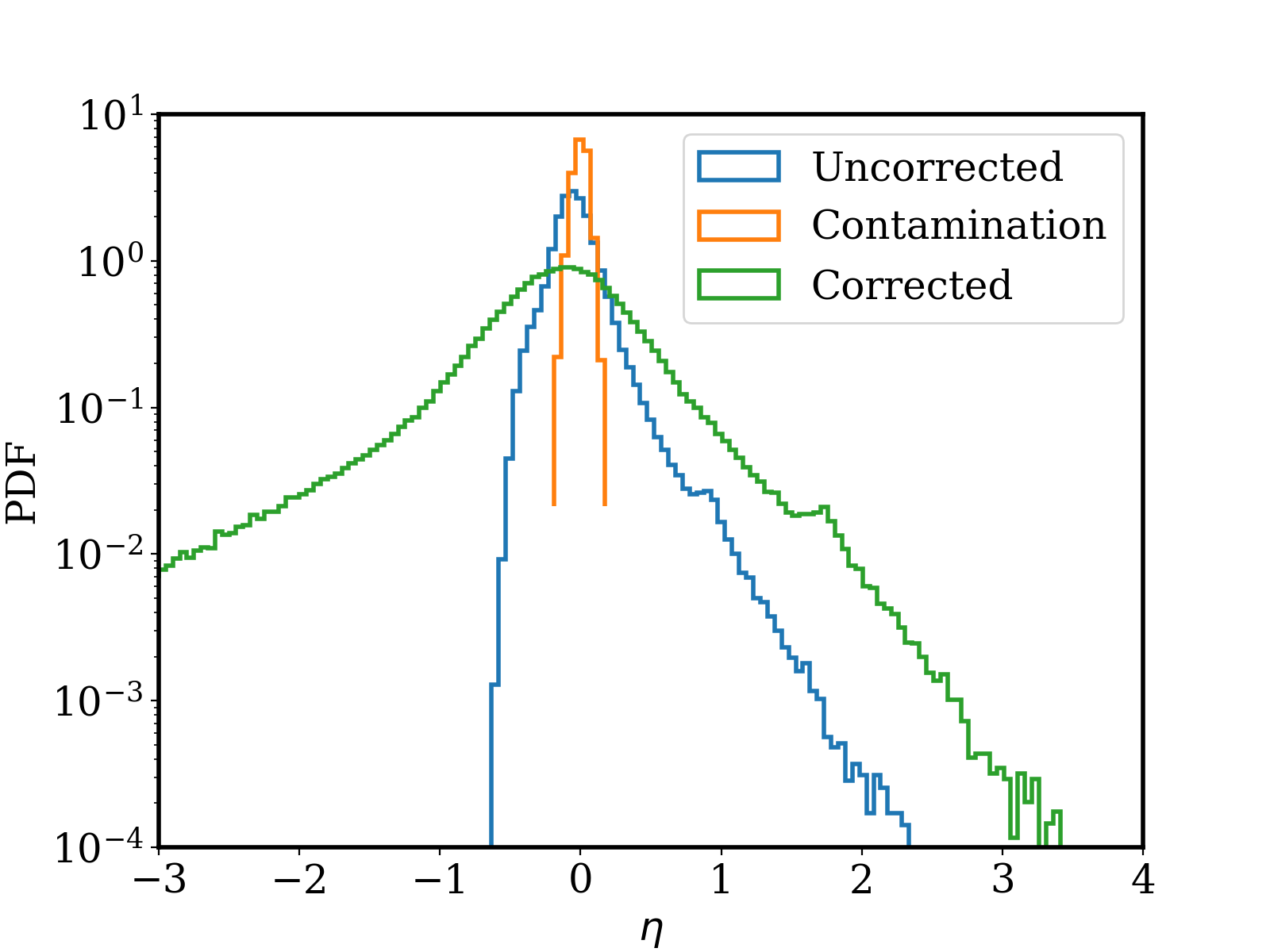}
\caption{A plot showing the column density distribution of the uncorrected column density map (in blue), the distribution in the contamination region (in orange), and the distribution for the corrected column density map (in green). The column density distributions are expressed in $\eta$ where $\eta = \ln{(\mathrm{N}_{H_2} / <\mathrm{N}_{H_2}>)}$.}
\label{fig::Correct}
\end{figure}  

\begin{figure*}
\centering
\includegraphics[width=0.8\linewidth]{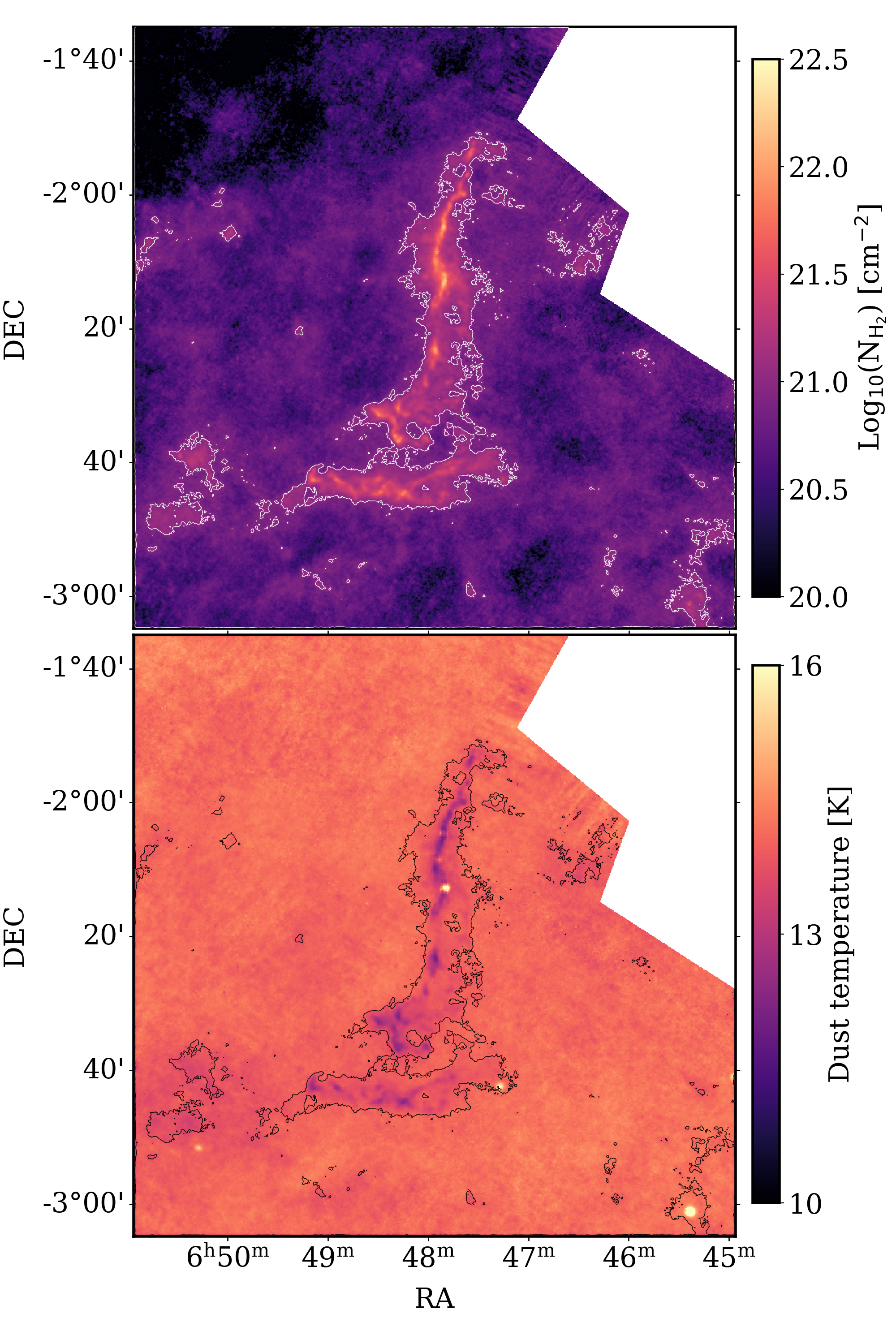}
\caption{(Top) The corrected column density map of G214.5-1.8 at 18.2'' resolution. (Bottom) The dust temperature map of G214.5-1.8 at 36.3'' resolution. The contours in both panels shows where the visual extinction is equal to 1 magnitude, i.e. $0.94 \times 10^{21}$ cm$^{-2}$.}
\label{fig::CDTEMP}
\end{figure*}  

\section{Results}\label{SEC:RES}%

\subsection{Column density and dust temperature maps}\label{SSEC:CDTEMP}%

We use the 160 $\mu$m, 250 $\mu$m, 350 $\mu$m and 500 $\mu$m Herschel images to produce `super-resolution' (18.2'') H$_2$ column density map and a standard (36.3'') resolution dust temperature map using the procedure described in \citet{Pal13}. We outline it here for completeness.

Assuming the dust emission to be optically thin in the wavelengths considered and that a single temperature may be used to describe it, one may fit a modified blackbody function to each pixel independent of each other to determine a column density, $\Sigma$, and a dust temperature, $T_d$. This modified blackbody function takes the form:
\begin{equation}
I_{\lambda}(\Sigma, T_d) = \kappa(\lambda) \; \Sigma \; \left(\frac{2 h c^2}{\lambda^5} \frac{1}{e^{hc / \lambda k_B T_d} - 1} \right),
\label{eq::MBB}
\end{equation}
where $h$ is Planck's constant, $k_B$ is the Boltzmann constant, $c$ is the speed of light and $I_{\lambda}$ is the surface brightness at wavelength $\lambda$. We use a mean molecular mass of 2.8 to convert from the total column density, $\Sigma$, to the column density of H$_2$ molecules, $\mathrm{N}_{H_2}$. The first term, $\kappa(\lambda)$, is the dust opacity per unit mass\footnote{Here a gas-to-dust ratio of 100 has been assumed and $\kappa$ is the dust opacity per unit total gas and dust mass.} as a function of wavelength. We adopt the power-law approximation:
\begin{equation}
\kappa(\lambda) = \kappa_{300} \left(\frac{300 \mathrm{\mu m}}{\lambda}\right)^{\beta},
\end{equation}
where $\kappa_{300}$ is the opacity per unit mass at 300 $\mu$m and $\beta$ is the dust emissivity index. We set $\kappa_{300} = 0.1$ cm$^2$/g and $\beta = 2$ \citep{Hil83}. 

To fit the modified black body equation we use a bounded least-squared fitting method. Each data point involved in the fit has an error. The error has two parts: the absolute calibration error and the map noise error. The absolute calibration error at 160 $\mu$m is given as 20$\%$ of $I_{160}$ and for the SPIRE bands it is given as 10$\%$ of $I_{250}$, $I_{350}$, and $I_{500}$ respectively \citep{Kon15}. The map noise error is estimated as the standard deviation of the intensity in that band from a small region in the top left of the considered map (100 x 100 pixels) as this region shows little/uniform emission. The total error is calculated by adding these two errors in quadrature to produce a conservative estimate of the total error.

To produce the `super-resolution' column density map, $\Sigma_s$, one first produces three column density maps, $\Sigma_{500}$, $\Sigma_{350}$ and $\Sigma_{250}$, at the angular resolution of their corresponding Herschel band, i.e. 36.3'', 24.9'' and 18.2''. The column density map $\Sigma_{500}$ is made by first smoothing the 160, 250 and 350 $\mu$m maps to the resolution of the 500 $\mu$m band. We fit the 4-band data with a modified black body as described above. The column density map $\Sigma_{350}$ is produced in a similar manner, but the 500 $\mu$m data is excluded and only the 160 and 250 $\mu$m image maps are smoothed to the resolution of the 350 $\mu$m band. This 3-band data is then fit with a modified black body. The column density map $\Sigma_{250}$ is produced by smoothing the 160 $\mu$m image map to the resolution of the 250 $\mu$m band and using these two bands to fit the modified black body function. Due to the limited information, the temperature is not a free parameter in the fitting but is given by the flux ratio $F_{250}/F_{160}$. This is achieved by using the bisector method to determine the root of the equation:
\begin{equation}
\frac{F_{250}}{F_{150}} = \left( \frac{250 \, \mu m}{160 \, \mu m} \right)^{3+\beta} \left(\frac{\exp{(\lambda_T / 160 \, \mu m)} - 1}{\exp{(\lambda_T / 250 \, \mu m)} - 1} \right),
\end{equation}
where $\lambda_T = hc/k_B T$ \citep{She09}. The smoothing described above is achieved by convolving the image maps with a Gaussian kernel with a full width half maximum (FWHM) in arcseconds of $\sqrt{\theta_D^2 - \theta_B^2}$, where $\theta_D$ is the desired resolution and $\theta_B$ is the native resolution of the image map. 

Combining the three column density maps one can produce the `super-resolution' column density map, $\Sigma_s$, using the following equation:
\begin{equation}
\Sigma_s = \Sigma_{500} + \Sigma_{350} - \Sigma_{350} \circledast G_1 \\ + \Sigma_{250} - \Sigma_{250} \circledast G_2,
\end{equation}
where $\circledast$ denotes a convolution and, $G_1$ and $G_2$ are Gaussian kernels with FWHMs in arcseconds of $\sqrt{36.3^2 - 24.9^2}$ and $\sqrt{24.9^2 - 18.2^2}$ respectively. At this stage we re-grid the final column density map to a 6'' pixel size.

\begin{table*}
\centering
\begin{tabular}{|p{4cm}||p{3cm}|p{3cm}|p{3cm}|p{3cm}|}
\hline\hline
Quantity & Total map & G214.5-1.8 & Main filament & Head structure \\ \hline
Total mass & 40,800 M$_\odot$ & 16,200 M$_\odot$ & 11,900 M$_\odot$ & 4,280 M$_\odot$\\
Mass above A$_v$=1 & 11,100 M$_\odot$ & 8,660 M$_\odot$ & 5,930 M$_\odot$ & 2,730 M$_\odot$\\ 
$<\mathrm{\overline{N_{H_2}}}>$ above A$_v$=1 & $1.57 \times 10^{21}$ cm$^{-2}$ & $1.65 \times 10^{21}$ cm$^{-2}$ & $1.69 \times 10^{21}$ cm$^{-2}$ & $1.56 \times 10^{21}$ cm$^{-2}$\\
Area above A$_v$=1 & 317 pc$^2$ & 236 pc$^2$ & 157 pc$^2$ & 78 pc$^2$ \\
Median $T_d$ above A$_v$=1 & 13.8 K & 13.8 K & 13.8 K & 13.9 K \\\hline \hline
\end{tabular}
\centering
\caption{A table showing global properties of the total map, the G214.5-1.8 GMF, the main filament and the head structure (as defined in section \ref{SEC:G214}). All values are given to 3 significant figures.}
\label{tab::prop}
\end{table*}

\begin{figure}
\centering
\includegraphics[width=0.95\linewidth]{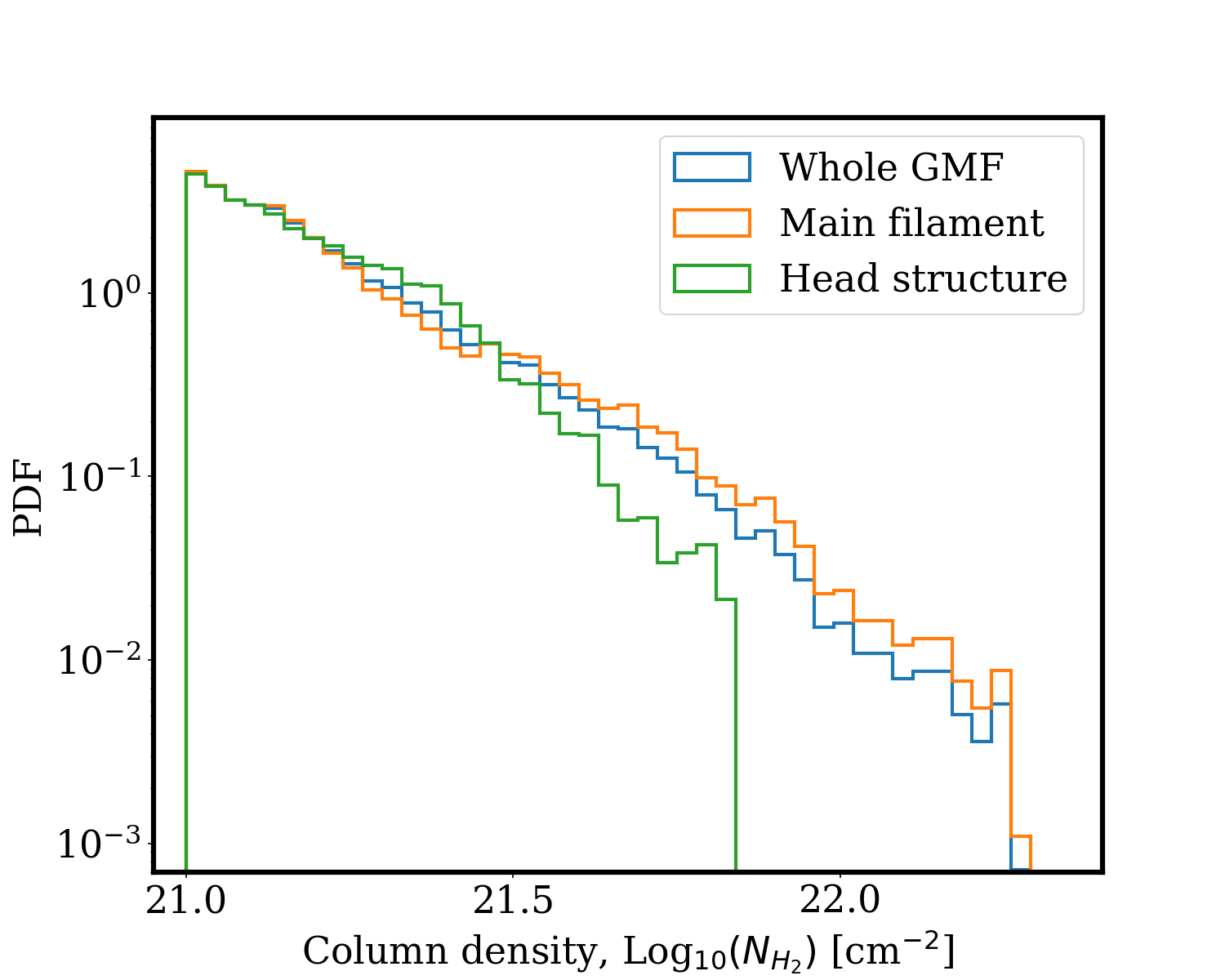}
\caption{The column density distributions of the whole G214.5 GMF (blue), the main filament section (orange), and the head structure (green) above a column density of $10^{21}$ cm$^{-2}$.}
\label{fig::NPDF}
\end{figure}  

The column density maps which result from the procedure as described suffer from line-of-sight contamination from unrelated dust emission. We follow the procedure outlined in \citet{Sch15} and \citet{Oss16} to estimate this contamination and correct for it. We first define a small rectangular region close to the border of the map (the same as that used to estimate map noise error, seen in figure \ref{fig::Bound}) which we assume is predominately dominated by the line-of-sight contamination. We determine a column density probability distribution (N-PDF) from the pixels within this region (plotted in figure \ref{fig::Correct}) and derive the peak and the width of this distribution. The peak of the contamination distribution is $1.2 \times 10^{21}$ cm$^{-2}$ compared to the peak of the uncorrected cloud distribution at $1.8 \times 10^{21}$ cm$^{-2}$. As this ratio is below one, and the width of the contamination is considerably narrower than that of the cloud distribution (as seen in figure \ref{fig::Correct}), the assumption of a simple uniform line-of-sight contamination is likely valid (as justified in \citet{Oss16}). We thus subtract $1.2 \times 10^{21}$ cm$^{-2}$, the peak of the contamination distribution, from every pixel of our uncorrected column density map to produce a `corrected' column density map. The column density distribution of the corrected map is also shown in figure \ref{fig::Correct}. Using multiple different small rectangular areas around the map edge yields contamination estimates of between $1.1 \times 10^{21}$ cm$^{-2}$ and $1.5 \times 10^{21}$ cm$^{-2}$. This variation adds an uncertainty in the final column density maps of only $\sim0.2 \times 10^{21}$ cm$^{-2}$, smaller than the possible error in Herschel column density maps, $\sim30\%$, due to variations in dust properties, potential dust temperature gradients along the line-of-sight, and assumptions about the opacity law \citep[see][for a discussion on this uncertainty]{Rus13}.

Finally, we manually trim the edges of the map to exclude the regions in which only the SPIRE band data is present, as well as the regions close to the map boundary where one sees clear mapping artefacts. The corrected column density map and the dust temperature map resulting from this process are shown in figure \ref{fig::CDTEMP}. We also show the contour relating to a visual extinction of 1 mag in figure \ref{fig::CDTEMP}; we use a conversion factor of 1 mag = $0.94 \times 10^{21}$ cm$^{-2}$ \citep{Boh78} here and throughout the paper. 

\subsection{Cloud properties}\label{SSEC:Prop}%

From figure \ref{fig::CDTEMP} one sees that the majority of gas above an $A_v$ of 1 mag is located within the G214.5 GMF, although there are a few small surrounding clumps. It is within the main filament and head structures which one sees the cold dust, as expected from well-shielded dense regions, and a number of small localised hot regions related to on-going star formation. 

Using the corrected column density map, and the assumption of a distance of 2.3 kpc, we determine a number of global properties of the cloud and the GMF. These are summarised in table \ref{tab::prop}. The majority of the cloud mass above an $A_v$ of 1 is located within the GMF ($\sim78\%$ of the total cloud). This is split between the main filament, containing $\sim68\%$ of G214.5's mass above an $A_v$ of 1, and the head structure, which contains the remaining $\sim32\%$. The average column density of the gas above an $A_v$ of 1 is comparable across both subregions, at $\sim 1.6 \times 10^{21}$ cm$^{-2}$ ($A_v \sim 1.7$ mag). However, defining dense gas as that above an $A_v$ of 7 mag one sees a stark difference. The main filament contains $\sim$374 M$_\odot$ of dense gas in comparison to the head structure containing only $\sim$3 M$_\odot$. This is further seen in figure \ref{fig::NPDF} which shows the N-PDFs of the two subregions above a column density of $10^{21}$ cm$^{-2}$ ($A_v \sim 1$ mag). Up to a column density of $\sim 3 \times 10^{21}$ cm$^{-2}$ ($A_v \sim 3$ mag), the head structure and the main filament show a comparable proportion of gas but above this quantity the distribution for the head structure rapidly falls while the main filament subregion extends up to $\sim 2 \times 10^{22}$ cm$^{-2}$ ($A_v \sim 20$ mag). Considering the whole GMF (it's N-PDF is also shown in figure \ref{fig::NPDF}), out of the 8,660 M$_\odot$ of gas above an $A_v$ of 1 mag only 377 M$_\odot$ is dense, corresponding to a dense gas mass fraction of 4.4\%. This paucity of dense gas in the GMF may be linked to the quiescence of G214.5. We do note that the dense gas mass fraction is a lower limit due to the beam dilution of the compact high column density structures; the 18.2'' column density map resolution corresponds to approximately 0.2 pc.

\begin{table*}
\centering
\begin{tabular}{|p{0.7cm}|p{1cm}|p{1cm}|p{1.2cm}|p{1cm}|p{1.3cm}|p{0.7cm}|p{0.7cm}|p{6cm}|}
\hline\hline

Source ID & RA [deg] & DEC [deg] & Peak $I_{70}$ [MJy/sr] & SNR & Total flux density [Jy] & $A_v$ [mag] & Peak? & Other known names \\ \hline
\multicolumn{3}{|l|}{Main filament}\\ \hline
S0 & 102.075 & -2.613 & 94.2 & 22.8 & 0.235 & 9.0 & Y & IRAS 06458-0233 \\
S1 & 102.085 & -2.601 & 236 & 57.3 & 0.692 & 8.0 & Y & IRAS 06458-0233 \\
S2 & 102.126 & -2.548 & 53.9 & 13.1 & 0.107 & 7.4 & Y & - \\
S3 & 102.079 & -2.532 & 47.9 & 11.6 & 0.067 & 6.9 & Y & - \\
S4 & 101.984 & -2.390 & 96.6 & 23.4 & 0.223 & 11.3 & Y & - \\
S5 & 101.984 & -2.383 & 51.2 & 12.4 & 0.111 & 11.3 & Y & - \\
S6 & 101.959 & -2.214 & 5628.9 & 1364.7 & 27.614 & 20.3 & Y & IRAS 06453-0209, S287-North$^{1}$ \\
S7 & 101.978 & -2.143 & 1635.0 & 396.4 & 6.234 & 10.5 & Y & IRAS 06453-0205 \\
S8 & 101.962 & -2.082 & 103.9 & 25.2 & 0.164 & 18.5 & Y & IRAS 06453-0201 \\
S9 & 101.963 & -2.077 & 858.0 & 208.0 & 2.823 & 18.5 & Y & IRAS 06453-0201 \\ \hline
\multicolumn{3}{|l|}{Head structure}\\ \hline
S10 & 102.091 & -2.751 & 25.2 & 6.1 & 0.037 & 2.8 & n & - \\ 
S11 & 102.063 & -2.736 & 224.0 & 54.3 & 0.718 & 6.8 & Y & - \\
S12 & 101.817 & -2.708 & 171.6 & 41.6 & 0.375 & 2.6 & Y & IRAS 06447-0239 \\
S13 & 101.822 & -2.707 & 142.8 & 34.6 & 0.592 & 1.5 & Y & IRAS 06447-0239 \\
S14 & 102.312 & -2.696 & 36.0 & 8.7 & 0.045 & 0.8 & n & - \\ \hline
\multicolumn{3}{|l|}{Surroundings}\\ \hline
S15 & 101.349 & -3.018 & 1783.1 & 432.3 & 22.154 & 7.1 & Y & IRAS 06428-0257 \\
S16 & 102.573 & -2.859 & 65.1 & 15.8 & 0.241 & 0.5 & n & PGC076120, a galaxy$^{2}$ \\
S17 & 101.766 & -2.762 & 57.4 & 13.9 & 0.120 & 0.8 & n & - \\
S18 & 102.617 & -2.738 & 68.0 & 16.5 & 0.162 & 1.3 & n & - \\
S19 & 101.267 & -2.686 & 56.9 & 13.8 & 0.129 & 0.6 & n & - \\
S20 & 101.490 & -2.654 & 31.6 & 7.7 & 0.052 & 0.6 & n & - \\
S21 & 102.621 & -2.653 & 73.9 & 17.9 & 0.153 & 2.4 & Y & - \\
S22 & 101.313 & -2.442 & 45.9 & 11.1 & 0.079 & 0.6 & n & MSX6C G214.4023-02.4899, M-type star $^{3}$ \\
S23 & 102.259 & -1.658 & 94.3 & 22.9 & 0.268 & 0.3 & n & IRAS 06465-0136, V377 Mon $^{4}$\\ \hline \hline
\end{tabular}
\centering
\caption{A table showing the 24 sources identified in the 70 $\mu$m map as described in section \ref{SSEC:70m}. The columns are: the source ID, the right ascension and declination (J2000) in degrees, the peak 70 $\mu$m intensity in the source in MJy/sr, the signal to noise ratio of the peak intensity, the total flux of the source in Jy, the peak $A_v$ in the source, if the source is associated with a column density peak, and other names and descriptors of the source. The references for the final column are: 1 - \citet{Bic03}, 2 - \citet{vdB15}, 3 - \citet{Mac10}, 4 - \citet{Ker94}. The sources are split into three groups: those located in the main filament (top), those located in the head structure (middle), and those located in the surroundings (bottom).}
\label{tab::70micron}
\end{table*}

Comparing G214.5 to the collection of clouds analysed in the same manner in \citet{Sch22}, one sees that G214.5 is very similar to the clouds they classify as `intermediate-mass star forming regions'. It has a mass above an $A_v$ of 1 greater than Mon R2 and Mon OB1 ($\sim$4,700 M$_{\odot}$), and slightly lower than that of NGC 2264 and Serpens (11,000-14,000 M$_{\odot}$). It also has a comparable mean column density and area above an $A_v$ of 1 to these clouds ($1.5-2.6 \times 10^{21}$ cm$^{-2}$, 60-270 pc$^2$). Despite the similarity in global properties, G214.5 is considerably more quiescent than these intermediate-mass star forming regions which are characterised by wide-spread low- and intermediate-mass star formation, as well as some high-mass star formation.

We note that we find a total mass for G214.5 of 16,000 M$_{\odot}$ which is nearly half that estimated by \citet{Yan19} using CO emission. Considering the errors associated with using the constant canonical X$_{\rm CO}$ factor to determine mass, these results are not in strong disagreement \citep{Bor21}. Interestingly, \citet{Bor21} note that young molecular clouds, i.e. before widespread star formation occurs, should have low X$_{\rm CO}$ factors ($\sim 1.0 \times 10^{20}$ cm$^{-2}$ (K km/s)$^{-1}$) which would bring the CO mass estimate into much better alignment with the dust emission estimate.

One may also study the global dust temperatures. The median dust temperature above an $A_v$ of 1 is 13.8 K and is nearly equal in both sub-regions of G214.5. When comparing to the catalogue of 45 GMFs presented in \citet{Zuc18}, one sees that G214.5 has a lower median dust temperature than any of those presented, making it the \textit{coldest giant molecular filament.} The low temperature may be related to the lower interstellar radiation field in the outer Galaxy and the low level of internal heating (seen in figure \ref{fig::CDTEMP}), indicative of an early evolutionary state.

\subsection{70 micron sources}\label{SSEC:70m}%

As mentioned in section \ref{SEC:G214}, G214.5 and the nearby Maddalena's cloud are quiescent compared to similar clouds in the inner galaxy. Here we use the Herschel data to compile a catalogue of 70 $\mu$m sources to determine the location and spread of active star formation in G214.5 and its immediate environment.

To identify sources we use the dendrogram package $\textsc{Astrodendro}$\footnote{http://www.dendrograms.org/}. To compute the dendrogram three parameters are needed: $\textsc{min\_value}$, the minimum value a pixel must have to be considered in the dendrogram; $\textsc{min\_delta}$, the minimum contrast needed to split a structure further; and $\textsc{min\_npix}$, the minimum number of pixels a structure must have. We set $\textsc{min\_value}$ to 5 times the noise level of the 70 $\mu$m map, $\textsc{min\_delta}$ to the noise level, and $\textsc{min\_npix}$ to 16. The choice of $\textsc{min\_npix}$ corresponds roughly to the area of a circle with a diameter the same as the FWHM of the 70 $\mu$m beam, 8.5''. The noise level of the map is determined by calculating the standard deviation of a 100 x 100 pixel region devoid of emission; it is found to be 4.1 MJy/sr. The exact choice of area does not greatly affect the noise level. We consider all leaves of the dendrogram (i.e. the structures which contain no substructure) to be 70 $\mu$m sources. Note that sources are only considered if they exist within the footprint of the column density map in figure \ref{fig::CDTEMP}. 

\begin{figure}
\centering
\includegraphics[width=0.99\linewidth]{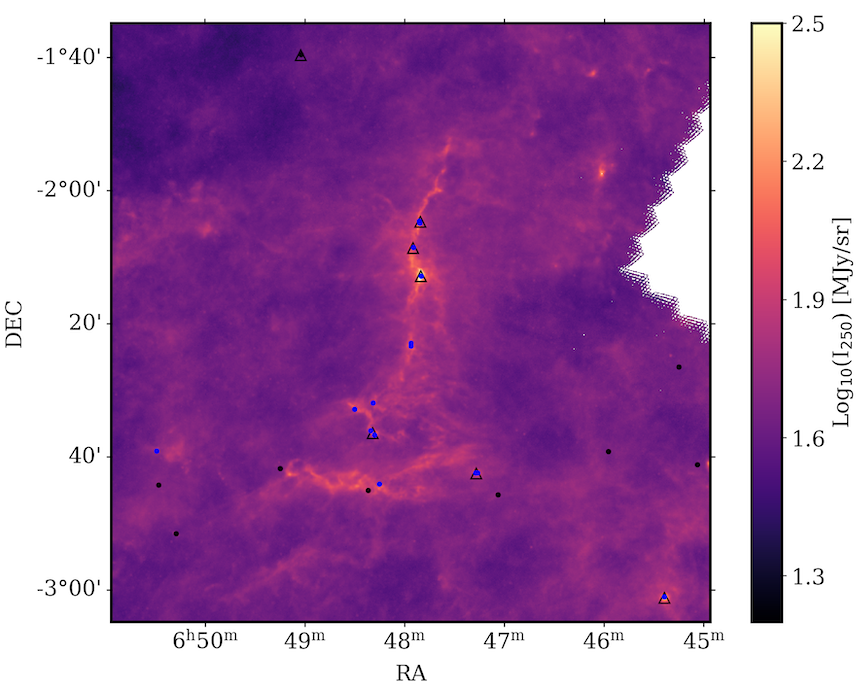}
\caption{A plot showing the location of the 24 70 $\mu$m sources overlaid on the 250 $\mu$m map. Sources which lie on, or close to, a column density peak are shown as blue circles and those which do not are shown as black circles. IRAS point sources in the region are shown as black triangles.}
\label{fig::70micron}
\end{figure}

\begin{figure}
\centering
\includegraphics[width=0.95\linewidth]{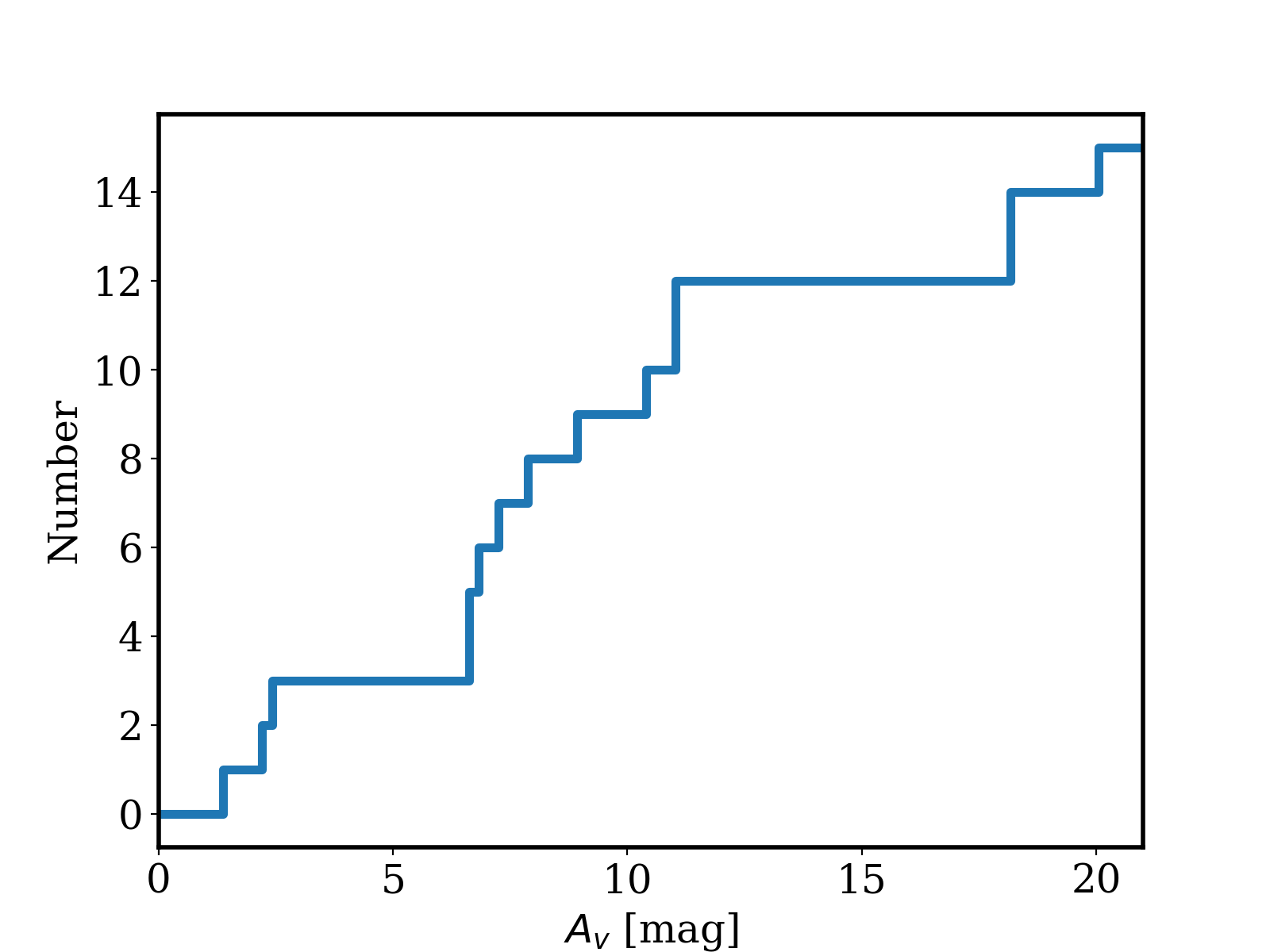}
\caption{A cumulative distribution of the peak $A_v$ of those 70 $\mu$m sources which lie on or close to a column density peak (i.e. the blue circles in figure \ref{fig::70micron}).}
\label{fig::AvDist}
\end{figure}

In total 24 sources are identified in the map; these are shown in figure \ref{fig::70micron} and their properties are summarised in table \ref{tab::70micron}. We separate the sources into two groups: those which are located within 18'' (the resolution of the column density map) of a column density peak and those which are not. This is done by visual inspection. There are 15 sources associated with column density peaks and 9 which are not. One sees that those 15 associated with column density peaks are predominately confined to the G214.5 GMF as expected and are most likely proto-stellar objects. All IRAS point sources \citep{IRASpoint} in the region (shown as black triangles) have corresponding 70 $\mu$m point sources, with three of them resolved into two distinct sources. Sources with no corresponding column density peaks are likely contaminants in the form of external galaxies or IR-bright stars. This is known to be the case for source S23 (co-incident with IRAS 06465-0136) which is a semi-regular variable star called V377 Mon \citep{Ker94}, for source S16 which is a massive external galaxy known as PGC076120 \citep{vdB15} and for source S22 which is a cool M-type star \citep{Mac10}. Investigating the 15 sources which are associated with column density peaks, one sees that they are associated with high column density gas, typically greater than an $A_v$ of 6 mag (see figure \ref{fig::AvDist}). Thus these sources are mostly deeply embedded, supporting their prestellar origin. 

Comparing the number of embedded sources in the G214.5 GMF, 13 of the 15 total embedded sources, to the widespread and large-scale proto-stellar activity in similar mass and size clouds such as Serpens, NGC 2264, Mon R2 and Mon OB1, one sees a clear deficit in G214.5's star formation \citep{Ray17,Fio21,Nony21}. Moreover, some of these clouds are forming high-mass stars, containing OB stars/clusters; however, source S6, which is coincident with IRAS 06453-0209 and is the brightest 70 $\mu$m source in G214.5, was found to show no sign of high mass star formation activity \citep{Lar99,Fur03,Sun07}, only two highly reddened sources in the K-band \citep{Lee96}. Thus we conclude that the G214.5 GMF is quiescent for its size and mass. If G214.5 is a young cloud and in a early phase of its evolution, then this quiescence, the scarcity of dense gas and the low average dust temperature shown in section \ref{SSEC:Prop} are naturally explained.

\section{Discussion}\label{SEC:DIS}%

\subsection{Clump properties}\label{SSEC:CLUMPS}%

\begin{figure*}
\centering
\includegraphics[width=0.8\linewidth]{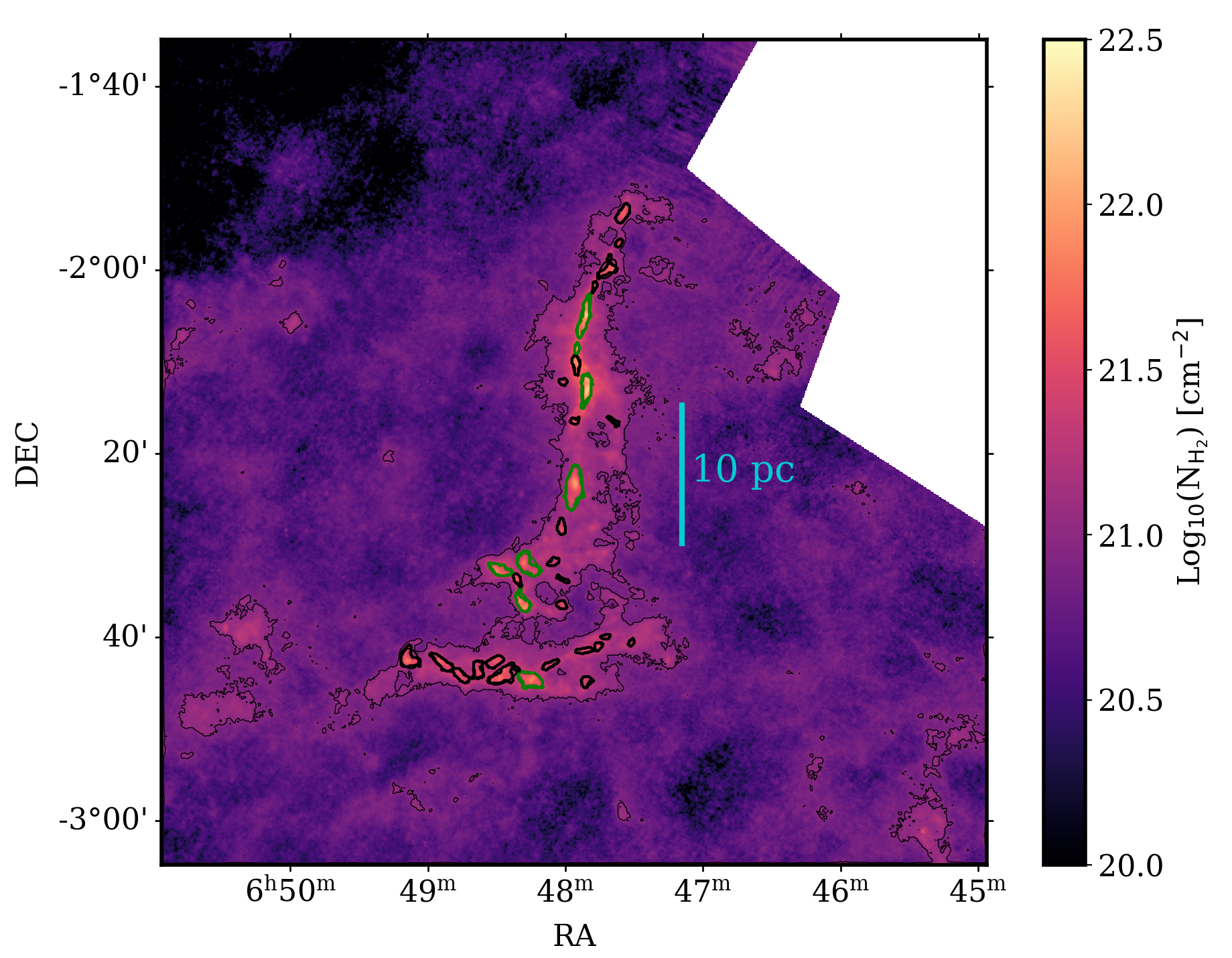}
\caption{A map showing the outlines of the 33 clumps identified overlaid on the column density map. Clumps shown in green harbour a 70 $\mu$m source inside their boundaries. Those which do not are shown in black. The grey contour shows the $A_v$ = 1 mag threshold for reference.}
\label{fig::clumps}
\end{figure*}  

\begin{table*}
\centering
\begin{tabular}{|p{1.2cm}|p{1.1cm}|p{1.2cm}|p{1.2cm}|p{1.2cm}|p{1.2cm}|p{1.2cm}|p{1.0cm}|p{1.2cm}|p{2cm}|}
\hline\hline

Clump ID & RA [deg] & DEC [deg] & Peak $A_v$ [mag] & Mass [M$_\odot$] & B$_{maj}$ [pc] & B$_{min}$ [pc] & $<T_d>$ [K] & L$_{\rm bol}$ $\;\;$ [L$_\odot$] & 70 $\mu$m assoc. \\ \hline
\multicolumn{3}{|l|}{Main filament}\\ \hline
C0 & 102.078 & -2.605 & 9.0 & 98.6 & 0.81 & 0.40 & 12.7 & 62.2 & S0,S1 \\
C1 & 102.007 & -2.608 & 3.8 & 20.4 & 0.40 & 0.30 & 13.0 & 9.1 & None \\
C2 & 102.087 & -2.562 & 4.9 & 28.5 & 0.53 & 0.22 & 12.8 & 10.9 & None \\
C3 & 102.004 & -2.563 & 2.3 & 7.9 & 0.54 & 0.13 & 13.4 & 4.6 & None \\
C4 & 102.069 & -2.535 & 6.9 & 106.5 & 1.04 & 0.57 & 12.9 & 67.0 & S3 \\
C5 & 102.119 & -2.545 & 7.4 & 88.0 & 0.87 & 0.37 & 12.8 & 50.1 & S2 \\
C6 & 102.021 & -2.531 & 2.5 & 16.1 & 0.46 & 0.26 & 13.3 & 8.9 & None \\
C7 & 102.007 & -2.468 & 4.2 & 29.9 & 0.61 & 0.29 & 13.3 & 15.3 &None \\
C8 & 101.984 & -2.394 & 11.3 & 214.1 & 1.57 & 0.60 & 13.0 & 139.0 & S4,S5 \\
C9 & 101.913 & -2.275 & 2.4 & 9.1 & 0.54 & 0.16 & 13.7 & 5.8 & None \\
C10 & 101.983 & -2.275 & 4.4 & 18.2 & 0.33 & 0.23 & 13.0 & 7.4 & None \\
C11 & 101.962 & -2.216 & 20.3 & 227.3 & 1.19 & 0.36 & 13.9 & 505.2 & S6 \\ 
C12 & 102.004 & -2.204 & 2.8 & 11.8 & 0.30 & 0.25 & 13.6 & 7.1 &None \\
C13 & 101.981 & -2.171 & 9.0 & 79.8 & 0.73 & 0.26 & 12.5 & 24.5 & None \\
C14 & 101.979 & -2.143 & 12.4 & 33.6 & 0.41 & 0.18 & 13.2 & 69.5 & S7 \\ 
C15 & 101.965 & -2.085 & 18.5 & 204.4 & 1.61 & 0.31 & 12.6 & 123.9 & S8,S9 \\ 
C16 & 101.947 & -2.030 & 5.8 & 19.6 & 0.42 & 0.15 & 12.5 & 6.1 & None \\
C17 & 101.922 & -1.999 & 5.9 & 70.5 & 0.81 & 0.46 & 13.0 & 28.9 & None \\
C18 & 101.903 & -1.952 & 4.4 & 13.4 & 0.34 & 0.22 & 13.2 & 6.4 & None \\
C19 & 101.895 & -1.898 & 4.9 & 55.3 & 0.78 & 0.36 & 13.1 & 25.0 & None \\ \hline
\multicolumn{3}{|l|}{Head structure}\\ \hline
C20 & 102.063 & -2.745 & 7.3 & 124.4 & 0.94 & 0.54 & 13.0 & 83.1 & S11 \\ 
C21 & 101.962 & -2.748 & 3.0 & 22.4 & 0.43 & 0.35 & 13.5 & 13.0 & None \\
C22 & 102.190 & -2.737 & 3.7 & 43.0 & 0.75 & 0.31 & 13.5 & 23.8 & None \\
C23 & 102.113 & -2.736 & 6.4 & 121.4 & 1.13 & 0.52 & 13.2 & 56.6 & None \\
C24 & 102.159 & -2.727 & 3.9 & 44.3 & 0.70 & 0.37 & 13.5 & 24.2 & None \\
C25 & 102.224 & -2.713 & 5.4 & 63.7 & 1.00 & 0.32 & 13.2 & 30.2 & None \\
C26 & 102.028 & -2.717 & 3.0 & 21.8 & 0.68 & 0.21 & 13.3 & 11.7 & None \\
C27 & 102.128 & -2.712 & 4.4 & 46.8 & 0.68 & 0.35 & 13.4 & 24.5 & None \\
C28 & 102.284 & -2.706 & 6.7 & 87.1 & 0.77 & 0.61 & 13.0 & 38.1 & None \\
C29 & 101.966 & -2.691 & 3.3 & 23.3 & 0.67 & 0.19 & 13.5 & 12.6 & None \\
C30 & 101.940 & -2.684 & 3.7 & 16.7 & 0.35 & 0.26 & 13.4 & 8.6 & None \\
C31 & 101.880 & -2.677 & 2.3 & 7.4 & 0.30 & 0.18 & 13.8 & 4.9 & None \\
C32 & 101.927 & -2.666 & 2.4 & 8.5 & 0.36 & 0.17 & 13.8 & 5.7 & None \\ \hline \hline
\end{tabular}
\centering
\caption{A table showing the properties of the 33 clumps identified in section \ref{SSEC:CLUMPS}. The columns are: the clump ID, the right ascension and declination (J2000) in degrees, the peak $A_v$ within the clump, the mass of the clump, the major and minor axes of the equivalent ellipse, the average dust temperature within the clump, the bolometric luminosity of the clump, and the associated 70 $\mu$m source (if there is one). The clumps are split into two groups, those located in the main filament (top) and those located in the head structure (bottom). }
\label{tab::clumps}
\end{table*}

Due to the fact the resolution of the column density map (18.2'') at 2.3 kpc is approximately 0.2 pc, we are unable to resolve individual core structures. However, slightly larger scale clumps, which could be fragmented into a number of cores, are resolved. To further investigate the quiescence of the G214.5 GMF we here study the properties of the numerous clumps which make up the GMF. 

To identify clumps we use the same dendrogram technique as used to identify 70 $\mu$m sources, considering all dendrogram leaves to be clumps. We construct the dendrogram using the corrected column density map produced in section \ref{SSEC:CDTEMP} using the parameters: $\textsc{min\_value} \, = \,1.86 \times 10^{21}$ cm$^{-2}$ ($A_v$ = 2 mag), $\textsc{min\_delta} \, = \, 2.35 \times 10^{20}$ cm$^{-2}$ ($A_v$ = 0.25 mag), and $\textsc{min\_npix} \, = \, 36$ (approximately four 18.2'' beam sizes). These values are chosen as they ensure that only dense and well resolved clumps are selected. The exact choice of $\textsc{min\_delta}$ has little result on the identified clumps. Increasing the value of $\textsc{min\_value}$ leads to lower mass clumps being excluded, while decreasing the value leads to numerous, unwanted small fluctuations being included; the value of $1.86 \times 10^{21}$ cm$^{-2}$ is thus a good compromise. Figure \ref{fig::clumps} shows the outline of the 33 clumps overlaid on the column density map, and table \ref{tab::clumps} summaries the properties of these clumps.

One can see that the clumps are solely located within the boundaries of the GMF, and are split between the main filament structure and the head structure. Each clump may be approximated as an ellipse with the same moments of inertia as its dendrogram structure, resulting in a major and minor axis. Note that these values are not deconvolved with the beam. The average radius, defined as the geometric average of the major and minor axes, range from 0.23 to 0.97 pc and has a median of 0.42 pc. The aspect ratio ranges from 1.2, roughly spherical, to 5.2, highly elongated, with a median of 2.1. Calculating the clump masses from the exact dendrogram structure one finds that it ranges from 7.4 to 227.3 M$_\odot$, with a median mass of 33.6 M$_\odot$. Thus the clump population consists of small low mass clumps as well as larger high mass clumps. The total mass in clumps is 1984 M$_\odot$, making up $\sim23$ \% of the mass of G214 above an $A_v$ of 1 mag. 

\begin{figure}
\centering
\includegraphics[width=0.98\linewidth]{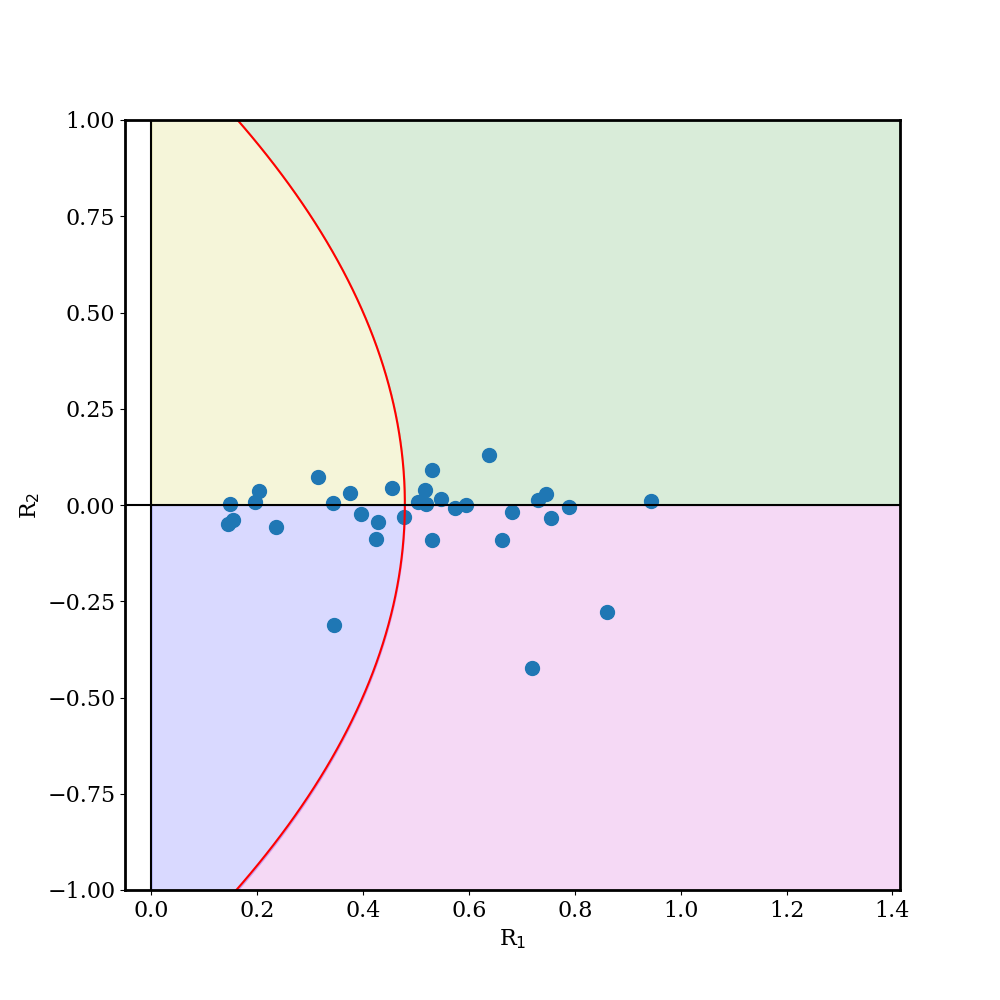}
\caption{An RJ-plot considering the column density maps of the 33 clumps identified in G214.5. The $R_1$ value increases with increasing elongation of a structure. Negative $R_2$ values are associated with central under-density, typically due to the curvature and asymmetry, while positive $R_2$ values indicate central over-density. The red, solid line is the boundary identified by \citet{Cla22} which separates quasi-circular objects, to the left of the line, from elongated structures, those to to the right. In conjunction with the $R_2=0$ line this produces the four classifications: (yellow) quasi-circular and centrally over-dense, (blue) quasi-circular and centrally under-dense, (green) elongated and centrally over-dense, and (magenta) elongated and centrally under-dense.}
\label{fig::RJ}
\end{figure}

We use the \textsc{RJ-plots} package from \citet{Cla22}\footnote{https://github.com/SeamusClarke/RJplots} to automatically classify the morphology of the clumps. Using the column density maps of the clumps, this provides two RJ-values, $R_1$ and $R_2$, which approximately separate elongation from central over-/under-density respectively. Figure \ref{fig::RJ} shows the RJ-plot resulting from considering the column density maps of the 33 clumps in G214.5. In total 18 of the 33 clumps are defined as being elongated (being in the green or magenta regions of the plot), and 17 of the 33 clumps are centrally over-dense (being in the yellow or green regions of the plot). In general the distribution is consistent with the clump morphology expected from filament fragmentation, as seen when comparing to the results of simulations presented in figure 8 of \citet{Cla22}, where clumps range from quasi-circular to highly elongated ($R_1>0.8$) and lie close to the $R_2=0$ line.

Cross referencing the embedded 70 $\mu$m sources found in the GMF with these clumps one finds that 8 of the 33 clumps harbour sources. We identify these clumps as protostellar as the presence of a 70 $\mu$m source is a reliable indicator of active star formation \citep{Dun08,Rag12}. Those without associated 70 $\mu$m sources we call starless clumps. Following \citet{Rag16}, we define the star-forming fraction (SFF) as the number of clumps harbouring a 70 $\mu$m source over the total number of clumps \footnote{We caution direct comparison between the SFF values quoted here and those in \citet{Rag16} due to the significantly different compact source extraction methods used.}. Seven of the eight protostellar clumps are located in the main filament.  The split of all clumps between the structures is 20 to 13, resulting in a difference of the star-forming fraction of 35.0\% to 7.6\%. Thus the main filament is the preferential region for star formation in G214.5 while the clumps in the head structure are distinctly non-active.

We determine the bolometric luminosity of the clumps in a manner similar to \citet{Elia17}. We use the Herschel flux density at 160, 250, 350 and 500 $\mu$m to fit a modified blackbody spectral energy distribution as done on a pixel by pixel basis in section \ref{SSEC:CDTEMP}. For starless clumps, the bolometric luminosity is equal to the integral of the fitted modified black body function from 1 $\mu$m to 10 cm. For protostellar clumps, the excess at shorter wavelengths must be taken into account; as such the bolometric luminosity is split into two: the infra-red luminosity ($< 160$ $\mu$m) and the sub-millimetre luminosity ($> 160$ $\mu$m). The sub-millimetre luminosity is determined by the integration of the modified blackbody function from 160 $\mu$m to 10 cm. For the infra-red luminosity we use the 70 $\mu$m flux density of the associated source, and take the 22 $\mu$m flux density as the sum of contributions of the point sources from the ALLWISE source catalogue \citep{Cur14} which lie within the clump. This is done as some clumps, such as C11, can be seen to harbour multiple WISE sources. To calculate the luminosity we linearly interpolate between the 22, 70 and 160 $\mu$m flux densities and integrate the area below this line. The resulting bolometric luminosities range from 4.6 to 505.2 L$_\odot$ with a median value of 23.8 L$_\odot$, leading to luminosity-mass ratios ranging from 0.31 to 2.22 L$_\odot$/M$_\odot$ and a median value of 0.55 L$_\odot$/M$_\odot$. Bolometric luminosities are also included in table \ref{tab::clumps}. 

\subsection{Clump statistics}\label{SSEC:CSTATS}%

\begin{figure}
\centering
\includegraphics[width=0.98\linewidth]{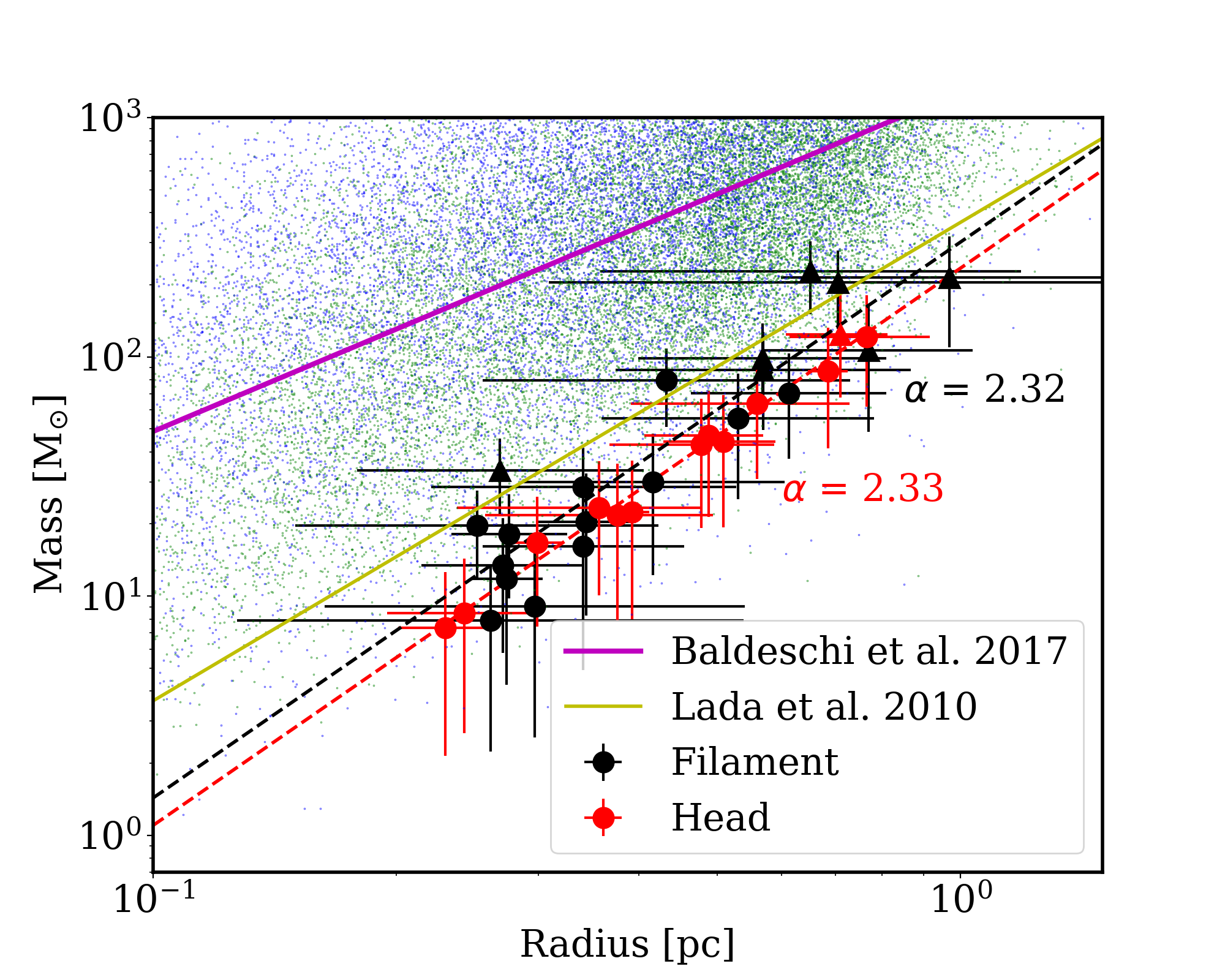}
\caption{A plot showing the mass-radius relationship of the main filament clumps (black) and the head clumps (red). Circles denote that the clump is starless, triangles show that the clump is protostellar. Power-law fits to the clump distributions are shown in dashed lines of their respective colour. Also shown as small dots are the protostellar (blue) and starless (green) clumps taken from the Hi-GAL survey \citep{Elia17} for reference, the high-mass star formation criterion of \citet{Bal17} is shown as a purple solid line, and the efficient star formation threshold of 116 M$_\odot$ pc$^{-2}$ proposed by \citet{Lada10} as a yellow solid line. The error-bars on the radius are given by the difference between the major and minor axes of the ellipse fitted to each clump; thus, more elongated clumps will have greater `errors'. The error-bars on the mass are calculated by adding in quadrature the uncertainty introduced by the background subtraction and the $30\%$ uncertainty from the column density estimation (as discussed in section \ref{SSEC:CDTEMP}). As such they should be considered highly conservative estimates of the error.}
\label{fig::massradius}
\end{figure}  

\begin{figure}
\centering
\includegraphics[width=0.98\linewidth]{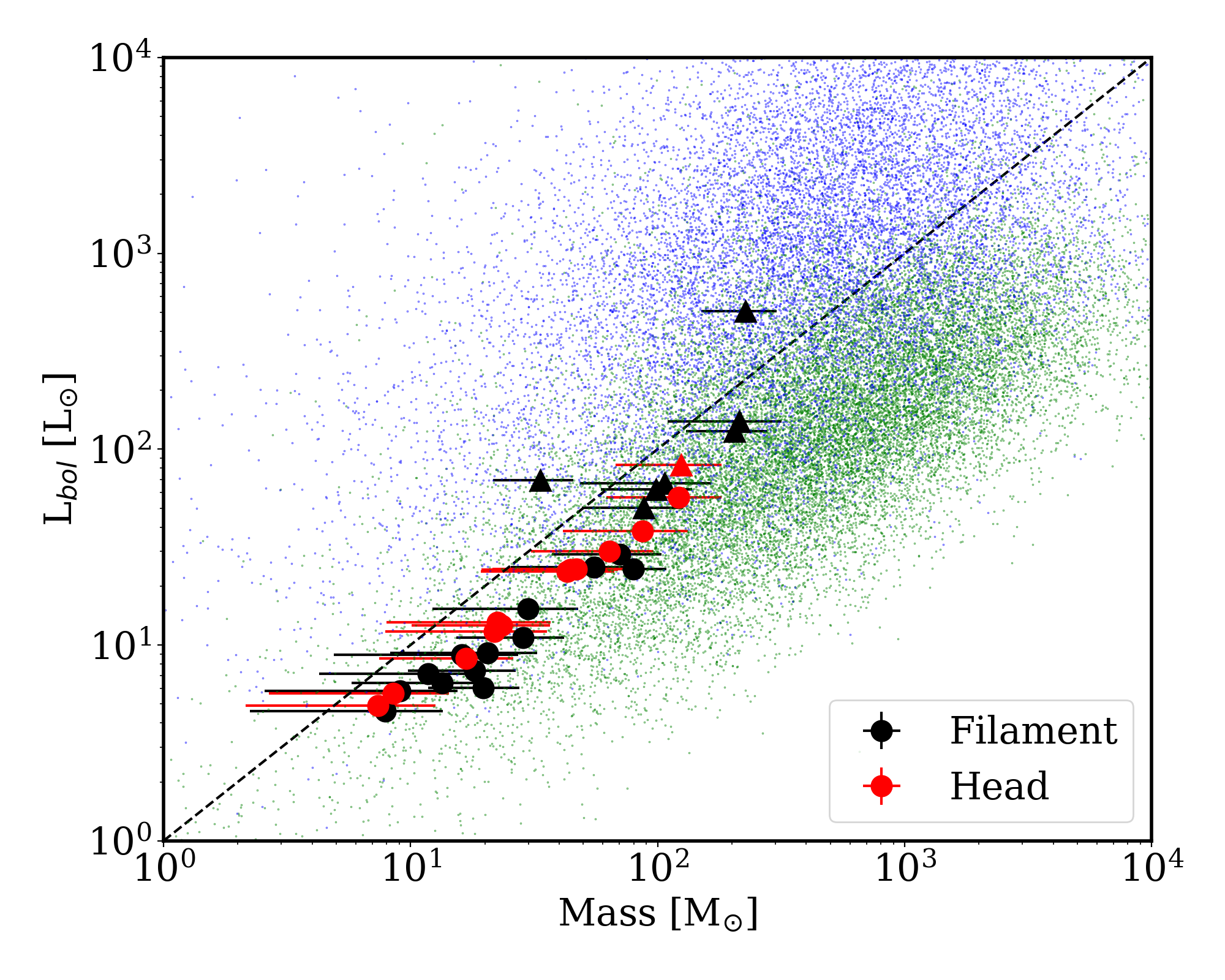}
\caption{A plot showing the mass-luminosity relationship of the main filament clumps (black) and the head clumps (red). Circles denote that the clump is starless, triangles show that the clump is protostellar. Also shown as small dots are the protostellar (blue) and starless (green) clumps taken from the Hi-GAL survey \citep{Elia17}. The dashed black line shows the luminosity to mass ratio of 1 L$_\odot$/M$_\odot$.The error-bars on the mass are calculated by adding in quadrature the uncertainty introduced by the background subtraction and the $30\%$ uncertainty from the column density estimation (as discussed in section \ref{SSEC:CDTEMP}). As such they should be considered highly conservative estimates of the error. The error-bars on the bolometric luminosity are determined by propagating the flux errors through the bolometric luminosity calculation described in section \ref{SSEC:CLUMPS} in a Monte-Carlo manner. They are of the order of $\sim1\%$.}
\label{fig::masslum}
\end{figure}  

\begin{figure}
\centering
\includegraphics[width=0.98\linewidth]{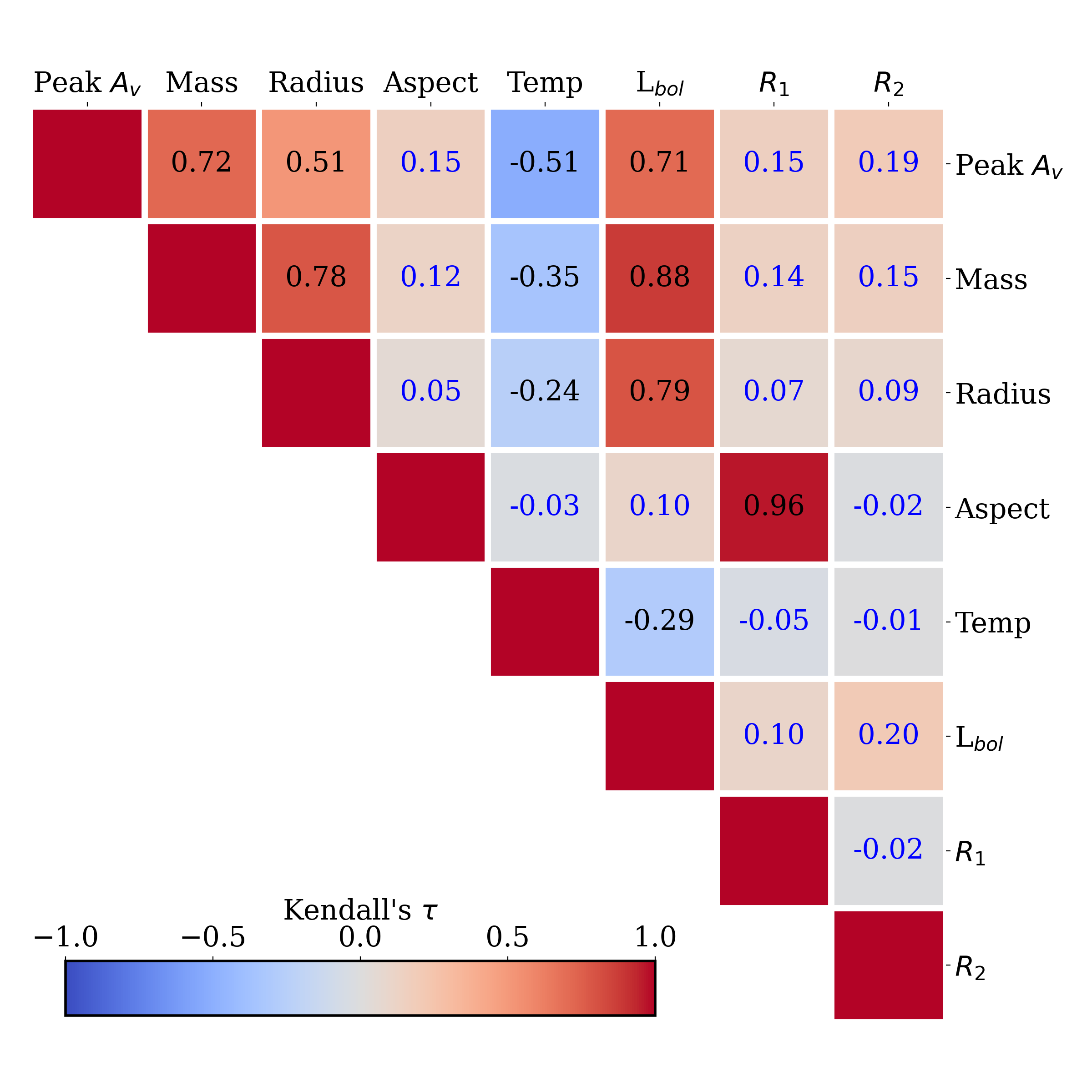}
\caption{A plot showing the correlation matrix considering all 33 clumps in G214.5 and the eight clump features: peak $A_v$, mass, radius, aspect ratio, average dust temperature, bolometric luminosity, and the RJ-values $R_1$ and $R_2$. Significant $\tau$ values (i.e. where $p<0.05$) are shown as black text and non-significant values (i.e. where $p>0.05$) are shown as blue text in each element of the matrix.}
\label{fig::corr}
\end{figure}  

As there is a clear difference in star formation activity between clumps found in the main filament and those in the head structure we use statistical techniques to investigate if there are differences in the two clump populations' properties. Moreover, this can be extended to compare between the G214.5 clumps and the wider clump population found as part of the Hi-GAL survey by \citet{Elia17, Elia21}. 

We begin by studying the clump mass-radius relationship (shown in figure \ref{fig::massradius}). One sees that there is a tight correlation between the quantities and that there is no clear difference between the filament and the head clumps, despite the vast difference in SFF. Using a least-squared method to fit a power-law relationship (M $\propto$ r$^{\alpha}$) between the two quantities results in nearly identical exponents of 2.32 and 2.33. Comparing the masses and radii of the G214.5 clumps to the \citet{Elia17,Elia21} sample, one sees that the G214.5 clumps are typically less massive for their size than the Hi-GAL clumps, indicative of lower average surface densities. However, this difference is due to the difference in clump identification techniques; the CuTEx method employed by \citet{Elia17,Elia21} returns a considerably smaller clump compared to the dendrogram method used here. This is shown in appendix \ref{APP:ELIA} where we show a comparison of the 16 co-incident clumps in the \citet{Elia21} sample and the catalogue compiled here. Excluding the background subtraction performed in section \ref{SSEC:CDTEMP} has only a small effect on the clump masses ($\sim20\%$) and does not affect any trends seen in clump properties.

Examining figure \ref{fig::massradius}, all 33 clumps lie below the high-mass star formation criterion from \citet{Bal17}, which is consistent with the lack of high-mass star formation tracers found when the most massive and brightest clump C11 (IRAS 06453-0209) was observed \citep{Lar99,Fur03,Sun07}. Additionally one sees that the majority of the clumps lie below the constant surface density line of 116 M$_\odot$ pc$^{-2}$ which is proposed by \citet{Lada10} as a threshold for efficient star formation. The lack of high surface density clumps is likely linked to the general quiescence of G214.5 despite its high mass. However, the median surface density of the filament and head clumps, 67 and 59 M$_\odot$ pc$^{-2}$ respectively, are very similar and cannot explain the difference in  clump star formation fraction (35.0\% to 7.6\%) between the two regions.

\begin{figure*}
\centering
\includegraphics[width=0.49\linewidth]{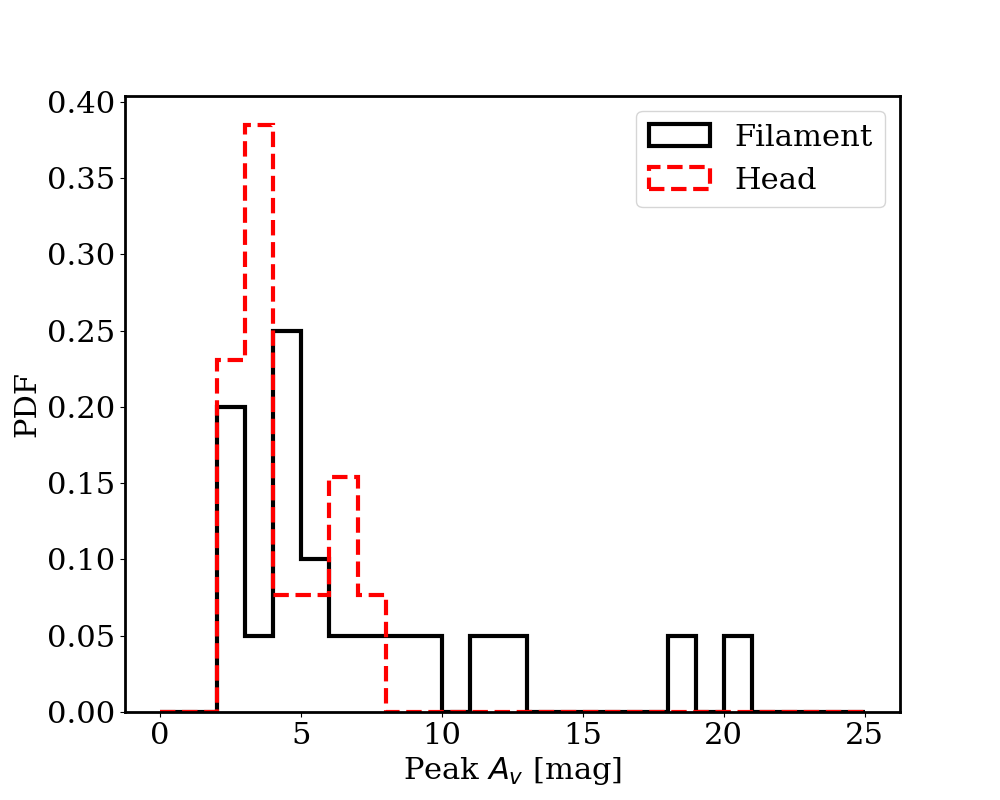}
\includegraphics[width=0.49\linewidth]{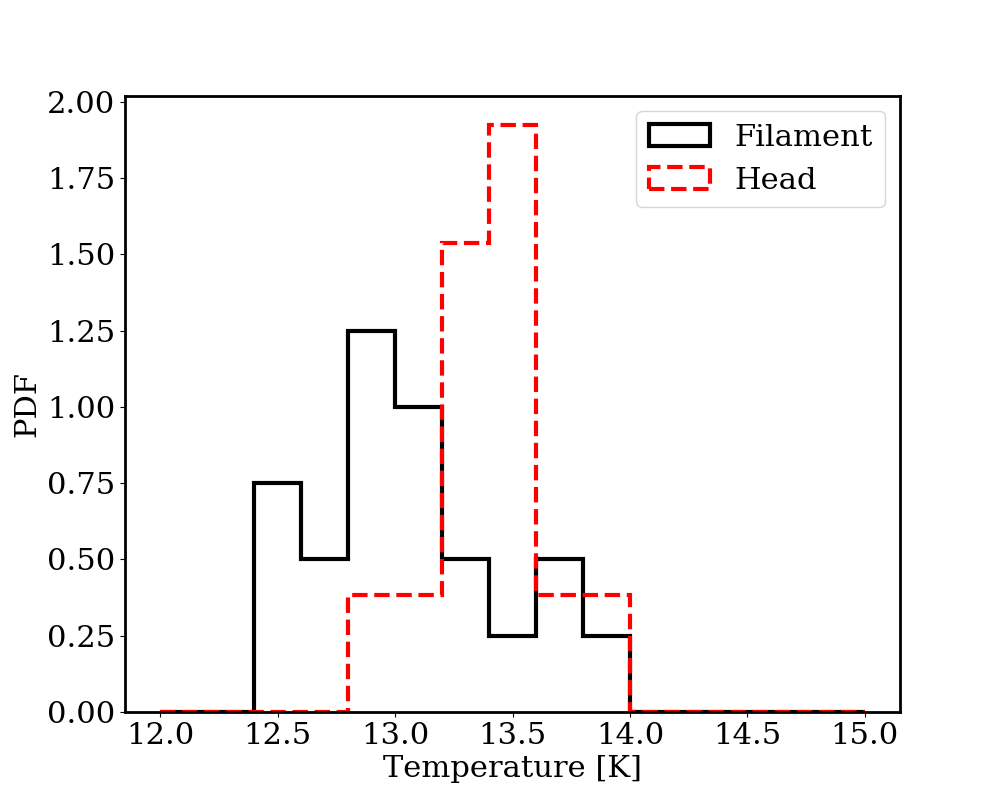}
\caption{Normalised histograms showing the distribution of (left) peak $A_v$ and (right) dust temperature of the clumps located in the main filament (black) and the head structure (red).}
\label{fig::avtemp}
\end{figure*}  

In figure \ref{fig::masslum} we show the mass-luminosity relationship of the G214.5 clumps, as well as the wider Hi-GAL sample. One sees little difference between the head and main filament clumps and a strong correlation between the two quantities. All 33 clumps are consistent with the distribution of Hi-GAL sample, with the majority of the clumps lying in the overlap region between the starless and protostellar distributions, or solely in the starless region. Only C11 and C14 lie in the clear protostellar region, likely due to hosting the two brightest 70 $\mu$m sources in the GMF, S6 and S7 respectively. These clumps have bolometric luminosity to mass ratios of $\sim 2.1$ L$_\odot$/M$_\odot$; the other 31 clumps have a ratio below 1 with a median value of 0.54 L$_\odot$/M$_\odot$. Considering the evolutionary tracks of \citet{Mol08}, the positions of the G214.5 clumps in the luminosity-mass plane are indicative of an early evolutionary state, consistent with the fact that G214.5 may be a young GMF.

To avoid investigating the correlation between each pair of clump features individually we use a correlation matrix to condense the information. We use the Kendall rank correlation test, quantified as Kendall's $\tau$, as a measure of the correlation due to its non-parametric nature and robustness \citep{Ken38}. We consider eight features of the clumps: the peak $A_v$, the mass, the radius, the aspect ratio, the average dust temperature, the bolometric luminosity, and the RJ-values $R_1$ and $R_2$. Figure \ref{fig::corr} shows the correlation matrix derived for all 33 clumps; significant $\tau$ values (i.e. where $p<0.05$) are shown as black text and non-significant values (i.e. where $p>0.05$) are shown as blue text. 

One can see that there exist strong and significant correlations between most pairs of clump features, except the aspect ratio, and the RJ-values $R_1$ and $R_2$. Thus, it appears that in G214.5, clump morphology plays little role in determining clump properties. The dust temperature is the only feature which exhibits negative correlations with other features. Comparing the correlation matrices of the clumps found in the head structure and the main filament yields no striking differences between the two clump populations. In general, the correlations shown are expected and are in agreement with those seen in the Hi-GAL data \footnote{Peak $A_v$, aspect ratio, $R_1$ and $R_2$ values are not included in the \citet{Elia17} sample and so could not be compared.}; however, the weak negative correlation between average dust temperature and bolometric luminosity is unexpected and does not agree with the Hi-GAL sample where a weak positive correlation is found. The weak negative correlation found is a combination of the very small range of dust temperatures ($\sim$ 12.5 to 14 K) but the large range of clump masses ($\sim$ 7 to 230 M$_\odot$), and the fact that the dust temperature is negatively correlated with the clump mass (due to the majority of clumps being starless) while the clump mass is very strongly correlated to the bolometric luminosity ($\tau = 0.88$). 

Despite the vast difference in the star formation fraction of the clumps located in the main filament and those located in the head structures there appear to be no clear differences in the populations when considering correlations. Thus, we use the two-sided Anderson-Darling (AD) and Mann-Whitney U (MW) null hypothesis tests to compare the 1D distributions of the clump properties. We use the same properties as used in the correlation matrix. Out of the six properties only the peak $A_v$ (MW: $p=0.04$) and dust temperature (AD: $p=0.008$ and MW: $p=0.005$) distributions show statistically significant differences between the clumps in the main filament and the head structure. These distributions are shown in figure \ref{fig::avtemp}. There exists a tail of high peak $A_v$ values in the filament clump distribution which is absent in the head clump distribution. Taking a limit of 5.5 mag (as this is $\sim$ 116 M$_\odot$ pc$^{-2}$, the efficient star formation threshold proposed by \citet{Lada10}), 50 $\%$ of filament clumps have a peak $A_v$ above this value but this is the case for only 23 $\%$ of the head clumps. As expected from the strong anti-correlation between temperature and peak $A_v$ ($\tau = -0.51$, figure \ref{fig::corr}) the filament clumps having higher peak $A_v$ values than the head clumps are also colder than them. The overlap in distributions is due predominately to the internal heating from protostellar sources in the filament clumps. 

Thus, one sees that while the clumps in the main filament are not considerably different to those located in the head structure, and also show similar relationships (i.e. figures \ref{fig::massradius}, \ref{fig::masslum}), that they are more deeply embedded resulting in higher peak column densities and colder dust. This leads to a larger proportion of filament clumps having peak column densities lying close to, or above, the efficient star formation regime (see figure \ref{fig::avtemp}) and it is this difference which results in the higher star formation fraction in the filament compared to the head structure.   

\subsection{The main filament}\label{SSEC:FIL}%

\begin{figure}
\centering
\includegraphics[width=0.98\linewidth]{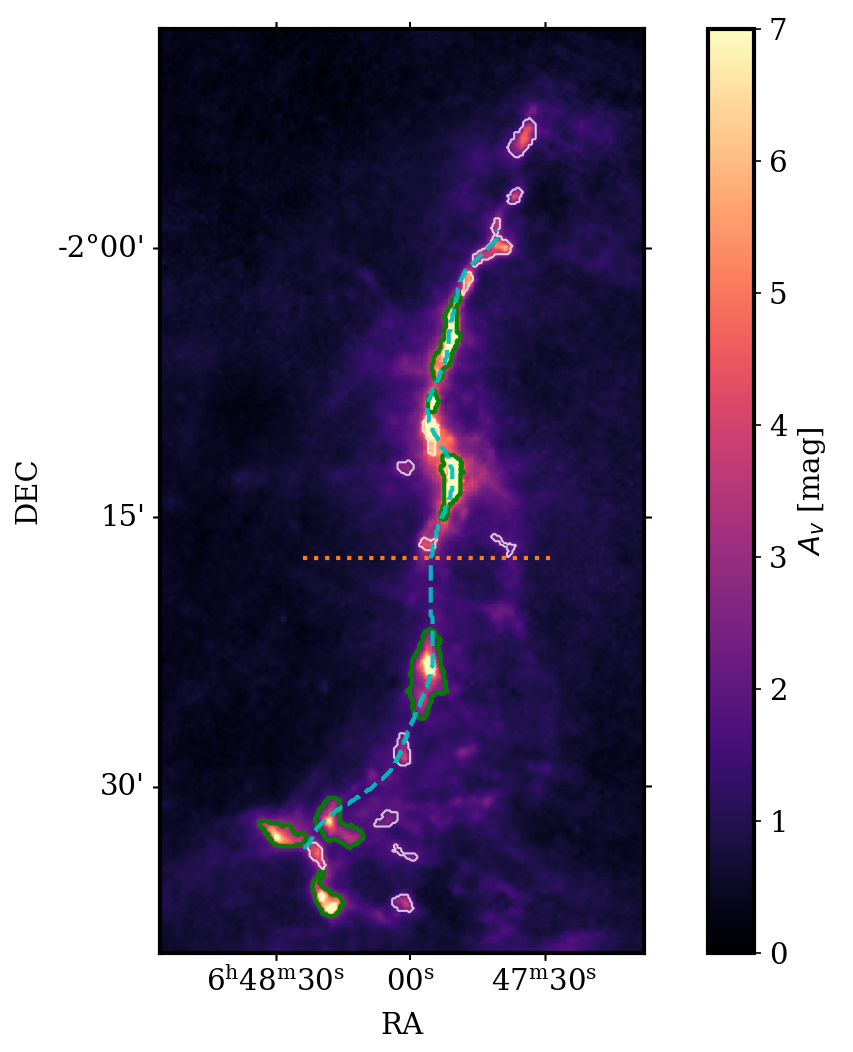}
\caption{A column density map showing the filament spine as a dashed, cyan line. Contours show the boundaries of the clumps located in the main filament (C0 to C19). Clumps shown in green harbour a 70 $\mu$m source inside their boundaries. Those which do not are shown in white. The horizontal, dotted, orange line denotes the separation between the upper and lower sections of the filament.}
\label{fig::spine}
\end{figure}  

Numerous giant molecular filaments have been studied by other authors \citep[e.g.][]{Rag14,Zuc18,Zha19} and their physical properties, such as length, line-mass and widths, have been reported. Due to the unique morphology of G214.5 we perform such an analysis only on the main filament region, excluding the head structure. We follow the methods presented in \citet{Cla19} and \citet{Cla20} to determine the filament properties and fragmentation.

We determine the main filament spine by following the same procedure laid out in section 4.2.2 of \citet{Cla20}: the column density map is convolved with a circular Gaussian kernel (here of size 10 pixels) to reduce the impact of sub-structure, a column density threshold (here 1.5 $\times 10^{21}$ cm$^{-2}$) is used to produce a binary image, a skeleton is built from the binary image using a medial axis transformation, and finally the skeleton is trimmed of unwanted side-branches, gaps are repaired and small manual corrections are made to ensure high column density ridges are followed. The exact choices of the smoothing length and column density threshold for this process do not greatly affect the spine found; these values are used as they reduce the number of side-branches and gaps in the skeleton. The resulting spine is shown in figure \ref{fig::spine} as a dashed cyan line. 

From figure \ref{fig::spine} one can see that the majority of the clumps located in the main filament lie close to or on the filament spine. Moreover, all of those which are protostellar lie on the spine, or close to the end in the cases of C0 and C4 in the south. The current star formation activity in G214.5 is thus confined to the dense ridge of the filament spine and is not widely distributed.

\begin{table}
\centering
\begin{tabular}{@{}*4l@{}}
\hline\hline
 & Whole  &  Upper  & Lower  \\ \hline
Length [pc] & 27.2 & 13.7 & 13.5 \\
Width [pc] & 0.87 & 0.71 & 2.16 \\
Aspect ratio & 31.2 & 19.3 & 6.25 \\
Mass [M$_{\odot}$] & 6570 & 3780 & 2790 \\
Line-mass [M$_{\odot}$/pc] & 242 & 277 & 207 \\
Spine $A_v$ [mag] & 4.07 & 6.12 & 2.40 \\
Spine $T_d$ [K] & 13.0 & 12.7 & 13.4 \\\hline
\end{tabular}
\centering
\caption{A table showing the global properties of the main filament when considering the whole filament, as well as when considering the upper and lower sections separately. The spine $A_v$ and $T_d$ quantities are median values along the filament spine defined in section \ref{SSEC:FIL}.}
\label{tab::fil}
\end{table}

\begin{figure*}
\centering
\includegraphics[width=0.9\linewidth]{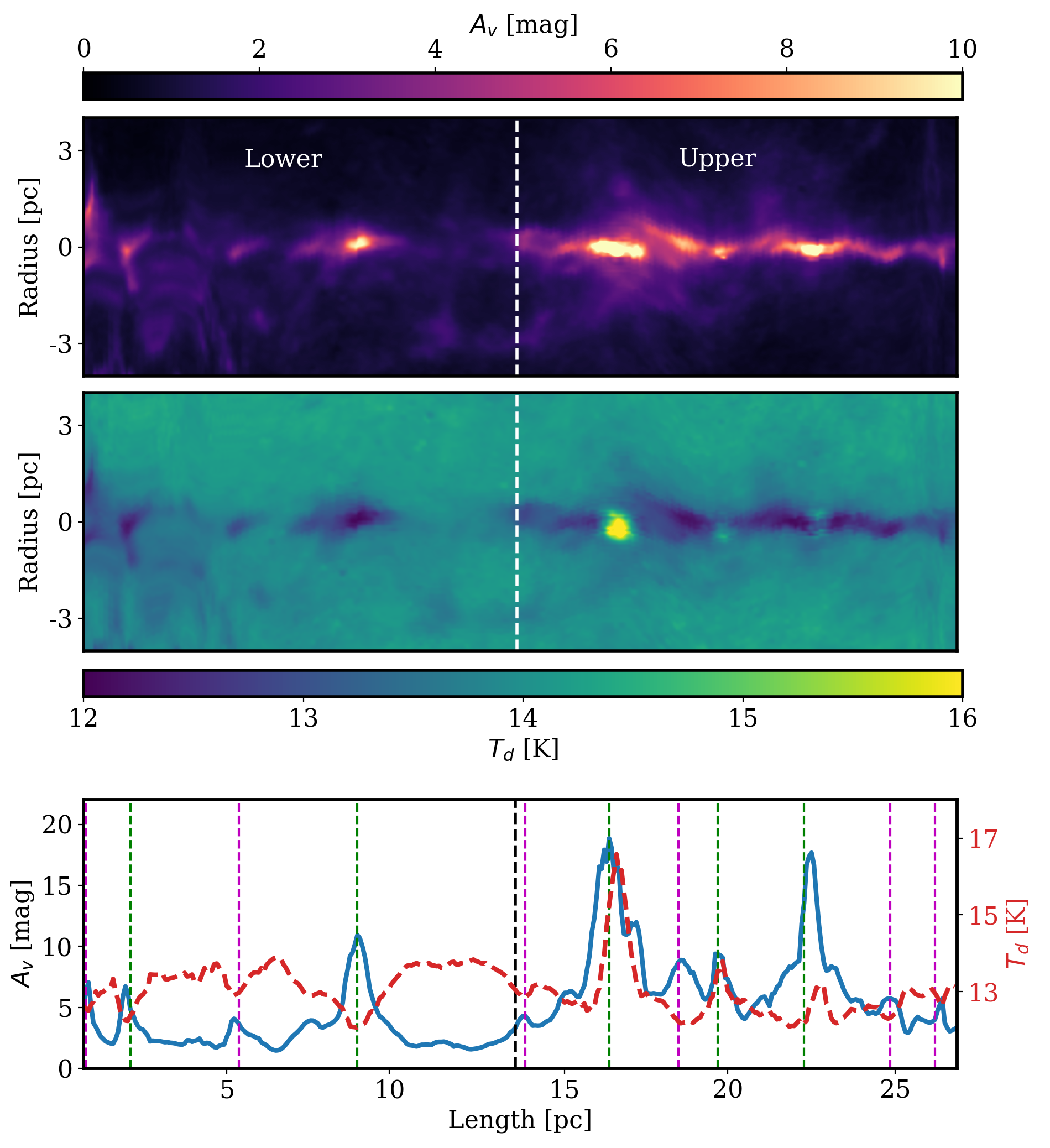}
\caption{(Top) The column density map of the straightened filament. (Middle) The dust temperature map of the straightened filament. (Bottom) The longitudinal profiles of the column density and dust temperature of the filament. The dashed vertical white/black line denotes the boundary between the lower and upper filament at l = 13.5 pc. The dashed purple and green vertical lines in the lower panel shows the mapped location of the starless and protostellar clumps respectively (discussed in section \ref{SSSEC:FRAG}).}
\label{fig::strfil}
\end{figure*}

As discussed in \citet{Cla19} one ought to straighten a filament, i.e. map it to the orthogonal radius-longitudinal distance space, before analysis to remove the complexity of the filament's curvature. We use the \textsc{Python} code \textsc{FragMent}\footnote{https://github.com/SeamusClarke/FragMent} for this purpose. The column density and dust temperature maps of the straightened filament is shown in figure \ref{fig::strfil}, as well as the corresponding longitudinal profiles. One can also see the clear presence of the cores located along the spine as local column density maxima, and those with embedded sources showing corresponding peaks in dust temperature.

When looking at figure \ref{fig::strfil} one can also see that the filament can be split into two distinct sections: a lower filament which is comprised of less dense, warmer and more flocculent material, and an upper filament which has a continuous, cold, dense spine and harbours the three brightest 70 $\mu$m sources in the GMF (S6, S7, S9) corresponding to localised high temperatures. Thus we consider these regions separately in our analysis using a boundary at l = 13.5 pc. Using this boundary we may quantify the difference between the two sections by looking at the median values of the column density and dust temperature along the spine. For the upper filament these values are 6.12 mag and 12.7 K, and for the lower filament they are 2.40 mag and 13.4 K.

The entire straightened filament has a length of 27.2 pc, resulting in similar lengths of the upper and lower sections, 13.7 and 13.5 pc respectively. The total mass within a radius of 4 pc of the entire filament spine is $\sim$ 6570 M$_{\odot}$, leading to an average line-mass of 242 M$_{\odot}$/pc. Considering the upper and lower sections separately, the total masses are 3780 M$_{\odot}$ and 2790 M$_{\odot}$ respectively with corresponding line-masses of 277 M$_{\odot}$/pc and 207 M$_{\odot}$/pc. The average line-mass of the filament is far in excess of the critical line-mass of $\sim 16$ M$_{\odot}$/pc for a thermally supported filament with gas temperature of 10 K \citep{Ost64}, making G214.5 highly supercritical. These global filament properties are summarised in table $\ref{tab::fil}$.

\subsubsection{The radial column density profile}\label{SSSEC:RADPRO}

\begin{figure}
\centering
\includegraphics[width=0.98\linewidth]{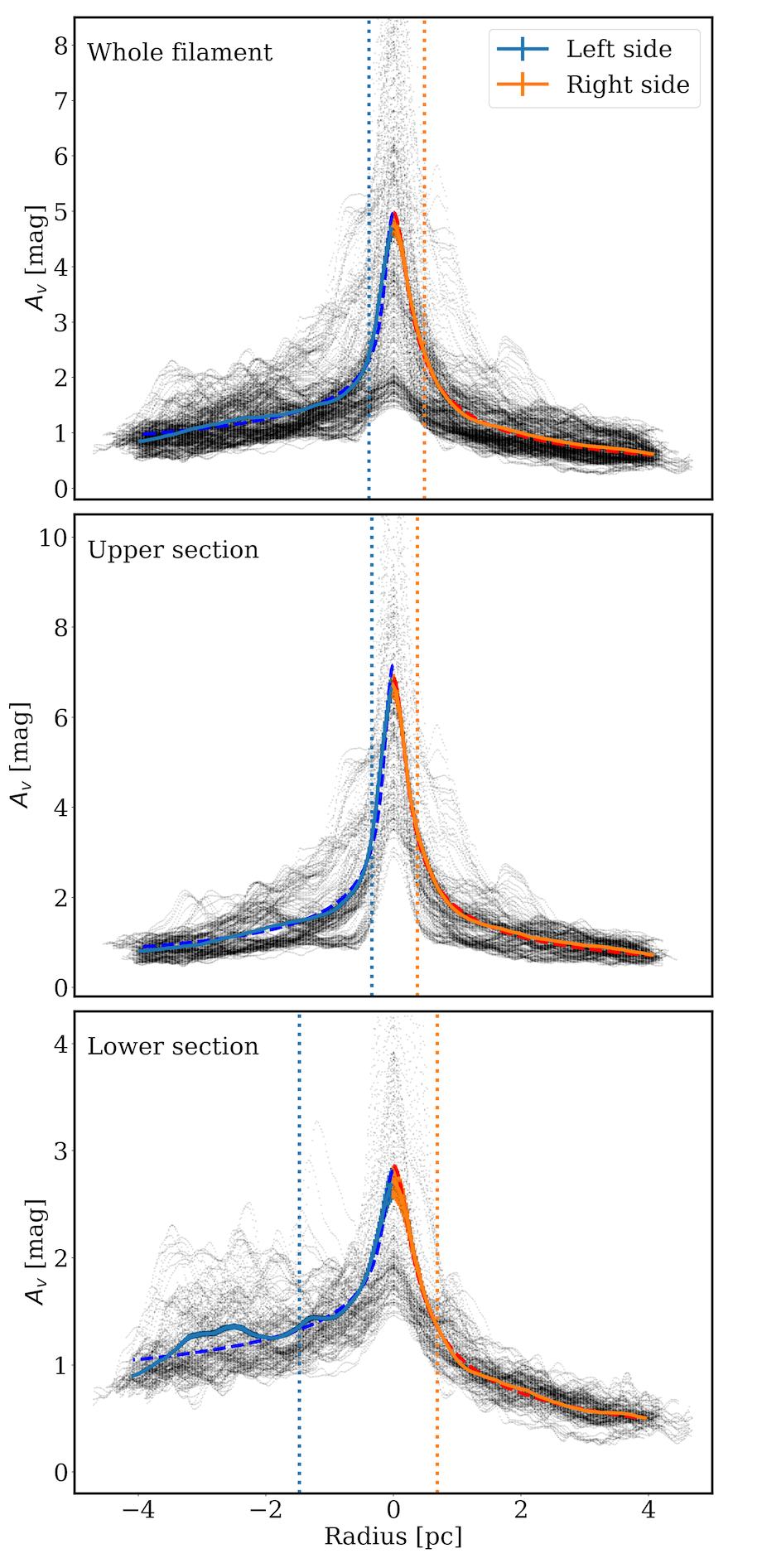}
\caption{The radial profiles of the (top) whole filament, (middle) the upper section, and (bottom) the lower section of the filament where the transparent black dots show the individual radial slices and the solid blue and orange lines show the mean radial profile constructed from those slices of the left and right sides of the profile respectively. The corresponding error bars shows the error on the mean. Also show as dashed blue and red lines are the Plummer fits to the inner 2 pc of each profile. The vertical blue and orange dotted lines show the observed HWHM of the right and left sides respectively.}
\label{fig::radpro}
\end{figure}

Another important aspect of the filament to study is the radial column density profile. The radial profile of the filament is shown in figure \ref{fig::radpro}. Each profile is centred such that the maximum column density occurs at $r=0$ pc as there is no guarantee that the spine will perfectly trace this point. From this centred radial profile we can define the left-hand side as $r<0$ pc and the right-hand side as $r>0$ pc, consistent with figure \ref{fig::strfil}. 

One can see from figure \ref{fig::radpro} that that there exists variation between the individual slices along the filament as well as considerable substructure to the sides of the filament spine. To reduce the complexity we can characterise the filament's radial profile by taking the mean of the individual slices (shown as solid lines in figure \ref{fig::radpro}). One can see that the mean profile of the whole filament is approximately symmetric, and this is also true when considering the upper section only. However, in the lower section there exists a clear asymmetry where the left-hand side is broader and shows a strong excess at larger radii compared to the right-hand side. This is not a result of averaging and can also be seen in the individual slices as well as the column density map of the straightened filament (figure \ref{fig::strfil}). Due to this asymmetry, when determining the widths we treat the left- and right-side of the profiles separately.

It is interesting to note that the asymmetric radial profiles seen in the lower section are qualitatively similar to those seen by \citet{Per12} and \citet{Zav20} who studied compressed filaments influenced by winds in the Pipe nebula and HII regions in RCW 120 respectively. However, here it is on the GMF scale rather than the $\sim 1$ pc length filaments observed previously. It is also interesting to note the recent work by \citet{Bon20b} investigating the Musca filament which links the asymmetric radial profiles of the filament to the influence of large-scale asymmetric HI flows.

As is common when studying filament profiles \citep[e.g.][]{Arz11}, we use a Plummer profile of the form:
\begin{equation}
\Sigma_{P}(r) = \frac{\Sigma_{o}}{\left(1 + (r/r_o)^2\right)^{\left(\frac{p-1}{2}\right)}},
\end{equation}
where $\Sigma_o$ is the central column density, $r_o$ is the flattening radius and $p$ is the radial profile's power-law slope when $r \gg r_o$. Note, we do not intend to claim that this is the true underlying radial profile of the filament, it is clearly more complicated as shown by the variety of individual slices, but it is an efficient method of characterising the filament's mean radial profile to readily compare to other examples. We also take into account the fact that the profile is beam-smeared by convolving the Plummer model, $\Sigma_{P}(r)$ ,with $g(r')$, a Gaussian function (FWHM = 0.203 pc, 18.2'' at 2.3 kpc). Written explicitly:
\begin{equation}
\Sigma(r) = \Sigma_{P}(r) * g(r').
\end{equation}
The fitting is achieved with a least-squared method and only the inner 2 pc region is used to avoid undue influence from side structures. The results are recorded in table \ref{tab::radprofit} and the best fits are shown in figure \ref{fig::radpro} as dashed lines; the Plummer models are seen to fit well to the observed column density profiles, even beyond the 2 pc fitting range. We do not perform background subtraction before fitting as correcting the column density map (section \ref{SSEC:CDTEMP}) removes unrelated line-of-sight dust; thus, all material in the radial profiles is material associated with the filament.

\begin{table}
\centering
\begin{tabular}{@{}*3l@{}}
\hline\hline
  Whole filament  & &  \\ \hline
 & Left side & Right side \\ \hline
$r_o$ [pc] & 0.009 $\pm$ 0.004 & 0.131 $\pm$ 0.009 \\
$p$ & 1.36 $\pm$ 0.01 & 1.64 $\pm$ 0.01 \\
$\Sigma_o$ [mag] & 9.04 $\pm$ 1.49 & 5.45 $\pm$ 0.16\\
$W$ [pc] & 0.380 & 0.490 \\ \hline \hline
 Upper section & & \\ \hline
& Left side & Right side \\ \hline
$r_o$ [pc] & 0.026 $\pm$ 0.007 & 0.097 $\pm$ 0.007 \\
$p$ & 1.49 $\pm$ 0.01 & 1.64 $\pm$ 0.01 \\
$\Sigma_o$ [mag] & 10.65 $\pm$ 1.08 & 7.90 $\pm$ 0.22\\
$W$ [pc] & 0.335 & 0.379 \\ \hline \hline
 Lower section & & \\ \hline
& Left side & Right side \\ \hline
$r_o$ [pc] & 0.027 $\pm$ 0.012 & 0.129 $\pm$ 0.017 \\
$p$ & 1.24 $\pm$ 0.01 & 1.53 $\pm$ 0.01 \\
$\Sigma_o$ [mag] & 3.43 $\pm$ 0.33 & 3.18 $\pm$ 0.17\\
$W$ [pc] & 1.472 & 0.691 \\ \hline \hline
\end{tabular}
\centering
\caption{A table showing the fitting parameters of the convolved Plummer profile model when applied to the left- and right-side of the whole filament, upper section only and lower section only mean column density profiles. The width reported, $W$, is the width at half-maximum of the corresponding column density profile.}
\label{tab::radprofit}
\end{table}

In all fits the resulting $r_o$ are found to be smaller than the beam-size, some as small as $0.01$ pc. These values should not be considered \textit{real}, but rather show that the inner region is highly beam-smeared and that the true narrowness of the filament is unresolved. Thus widths resulting from these Plummer fits, or from other forms of fitting in the inner region, would result in unreliable width estimation. As such we use the observed radius at which the profile reaches half the maximum of the column density profile to quantify the width (shown as vertical dotted lines in figure \ref{fig::radpro}). This should be considered as an upper estimate of the true underlying width. We see that the whole filament has a width of 0.87 pc, the upper section is slightly narrower, 0.71 pc, and that the lower section is considerably wider, 2.16 pc. Moreover, the asymmetry in the lower section is shown by the fact the the left-hand profile has a width over twice that the of the right-hand side, while the upper section is near symmetric. 

Unlike the flattening radius, $r_o$, the power-law exponents, $p$, resulting from the fits are unaffected by beam-dilution and are robust. These reveal an additional asymmetry in the profiles, that the left-hand profiles have shallower slopes compared to the right-hand profiles. This is apparent in both the upper and lower sections of the filament.

To further investigate the asymmetry between the two sides of the filament we use the asymmetry parameter developed by \citet{Per12}. The asymmetry parameter, $A_p$, is defined as:
\begin{equation}
A_p = \log{ \left( \frac{\int_{-r'}^{0} \Sigma_L(r) dr}{\int_{0}^{r'} \Sigma_R(r) dr} \right)},
\label{eq::ap}
\end{equation}
where $\Sigma_L$ is the observed column density profile on the left-hand side, $\Sigma_R$ is the same for the right-hand side, and $r'$ is the maximum radius of the shortest side of the profile to ensure the left- and right-hand profiles are integrated over the same radial extent. Here we have modified the original parameter by taking the logarithm so that the parameter is now symmetric around a value of 0 (meaning perfect symmetry between the left- and right-hand profiles). Considering the mean profiles of the lower and upper sections separately, the asymmetry parameter is 0.292 and 0.001 respectively. Thus the lower section of the filament has nearly twice as much mass on its left-hand side than on its right-hand side, while the upper section is close to perfectly symmetric in terms of mass. 

\begin{figure}
\centering
\includegraphics[width=0.98\linewidth]{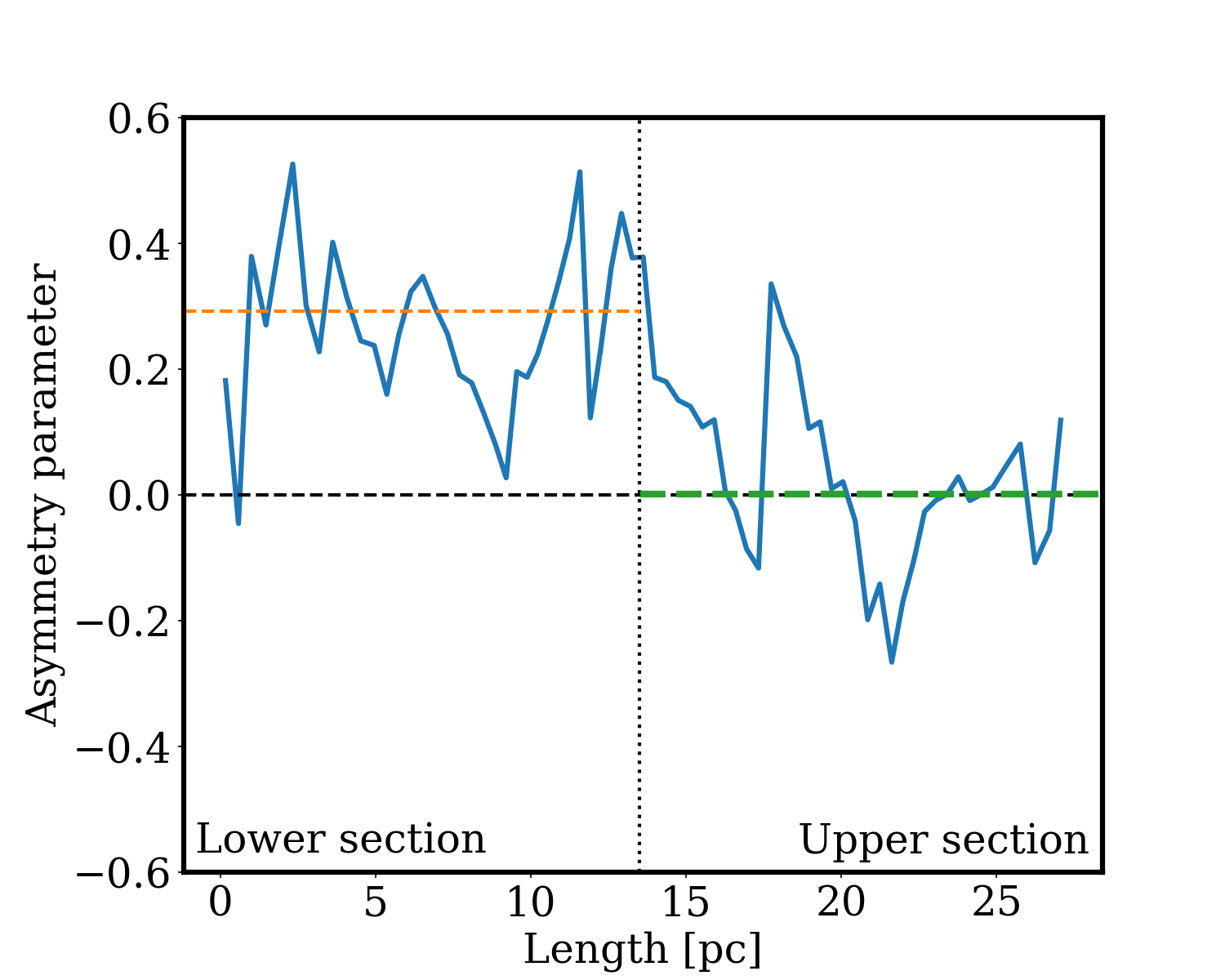}
\caption{The asymmetry parameter (equation \ref{eq::ap}) as a function of length along the filament. The vertical, dotted, black line at l=13.5 pc denotes the boundary between the lower and upper filament sections. The horizontal, dashed orange and green lines show the asymmetry parameter for the mean column density radial profile of the lower and upper sections respectively. The horizontal, dashed, black line is at $A_p = 0$, denoting perfect symmetry, to help orientate the reader.}
\label{fig::asym}
\end{figure}

To ensure that this is not due to averaging along the lengths of the filament sections we consider the asymmetry parameter locally. We calculate the mean radial profile of 5 consecutive individual slices to avoid excessive noise and determine the asymmetry parameter for this mean. As can be seen in figure \ref{fig::asym}, the lower section of the filament is consistently asymmetric in favour of the left-hand side and in some places reaches $A_p \sim 0.5$, i.e. the left-hand side has more than triple the mass than the right-hand side. The upper section oscillates between being left- and right-side heavy but rarely reaches the asymmetry seen in the lower section. The asymmetry parameters measured in the lower section are comparable to those found in the wind-compressed filaments in the Pipe nebula \citep[0.15 - 0.45, ][]{Per12}. Thus, from this analysis, we conclude that there is robust evidence that the lower section of G214.5 is asymmetric and likely being compressed from the east (this will be further discussed in section \ref{SEC:BUB}).

\subsubsection{Is there a characteristic fragmentation length-scale?}\label{SSSEC:FRAG}

\begin{figure*}
\centering
\includegraphics[width=0.48\linewidth]{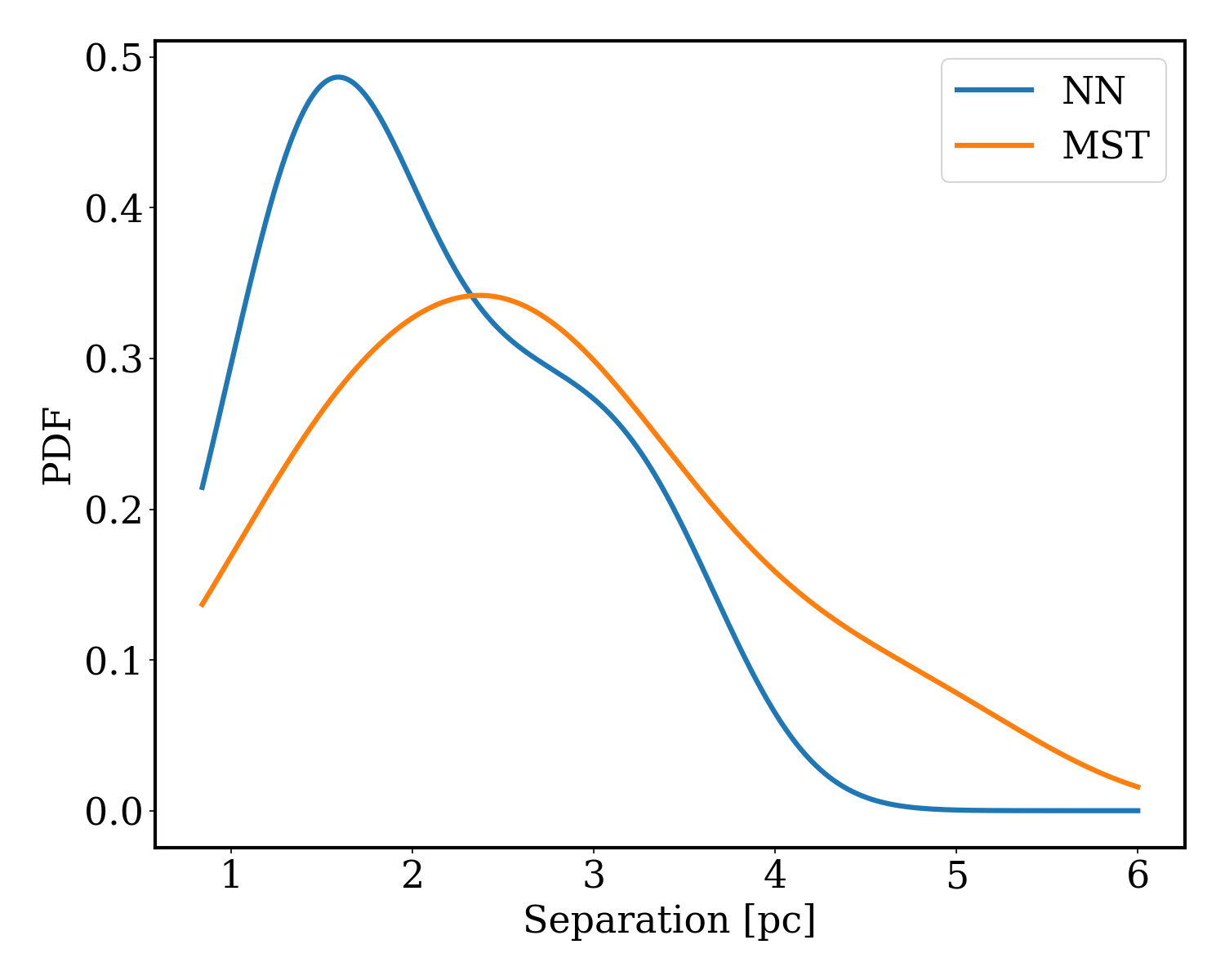}
\includegraphics[width=0.48\linewidth]{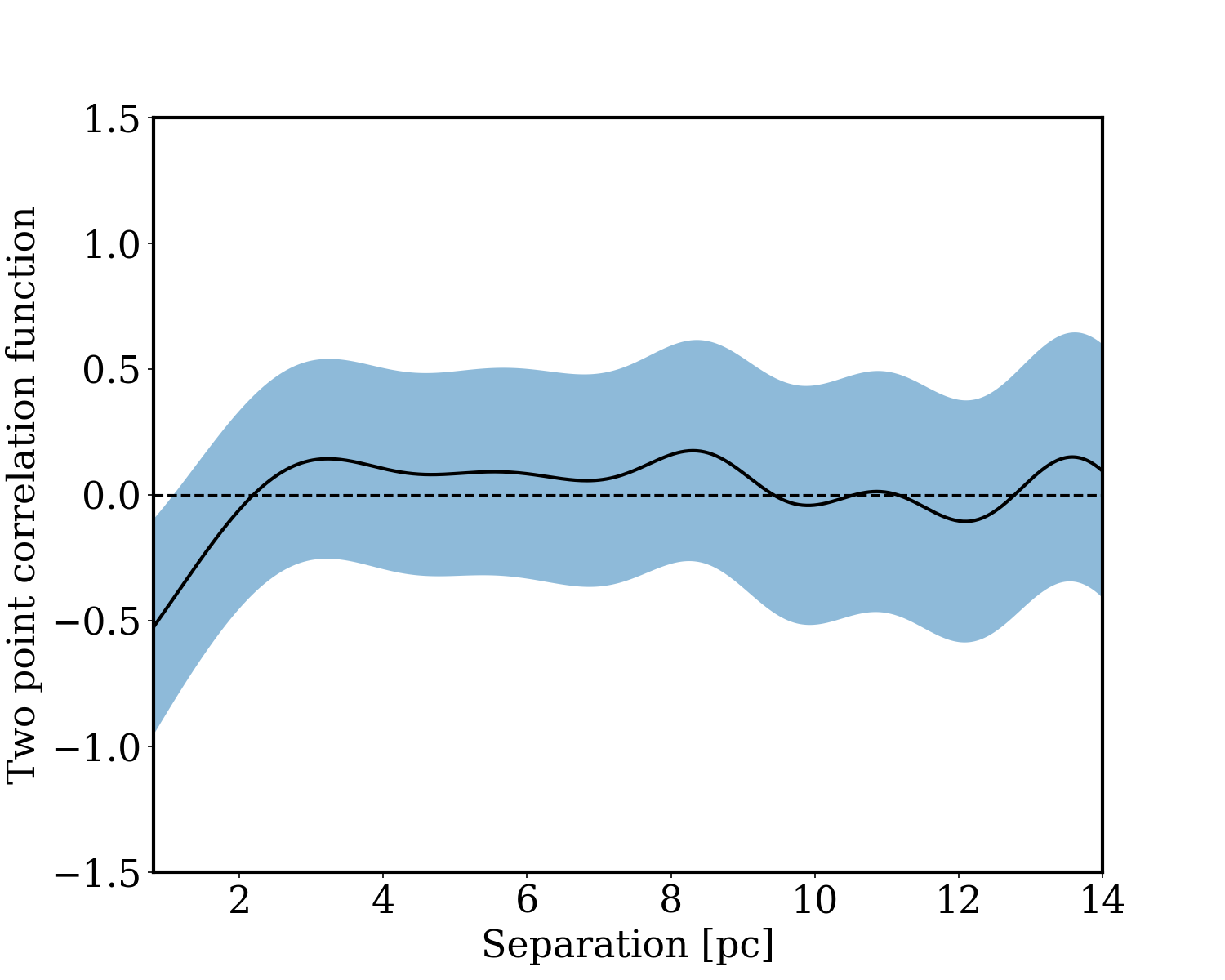}
\caption{(Left) The nearest neighbour (blue) and minimum spanning tree (orange) clump separation spacings. (Right) The two-point correlation function where the blue-shaded area is the 1$\sigma$ error. The horizontal dashed line at $y=0$ denotes no deviation from randomness and is added to aid the reader.}
\label{fig::charlength}
\end{figure*}  

\begin{figure}
\centering
\includegraphics[width=0.98\linewidth]{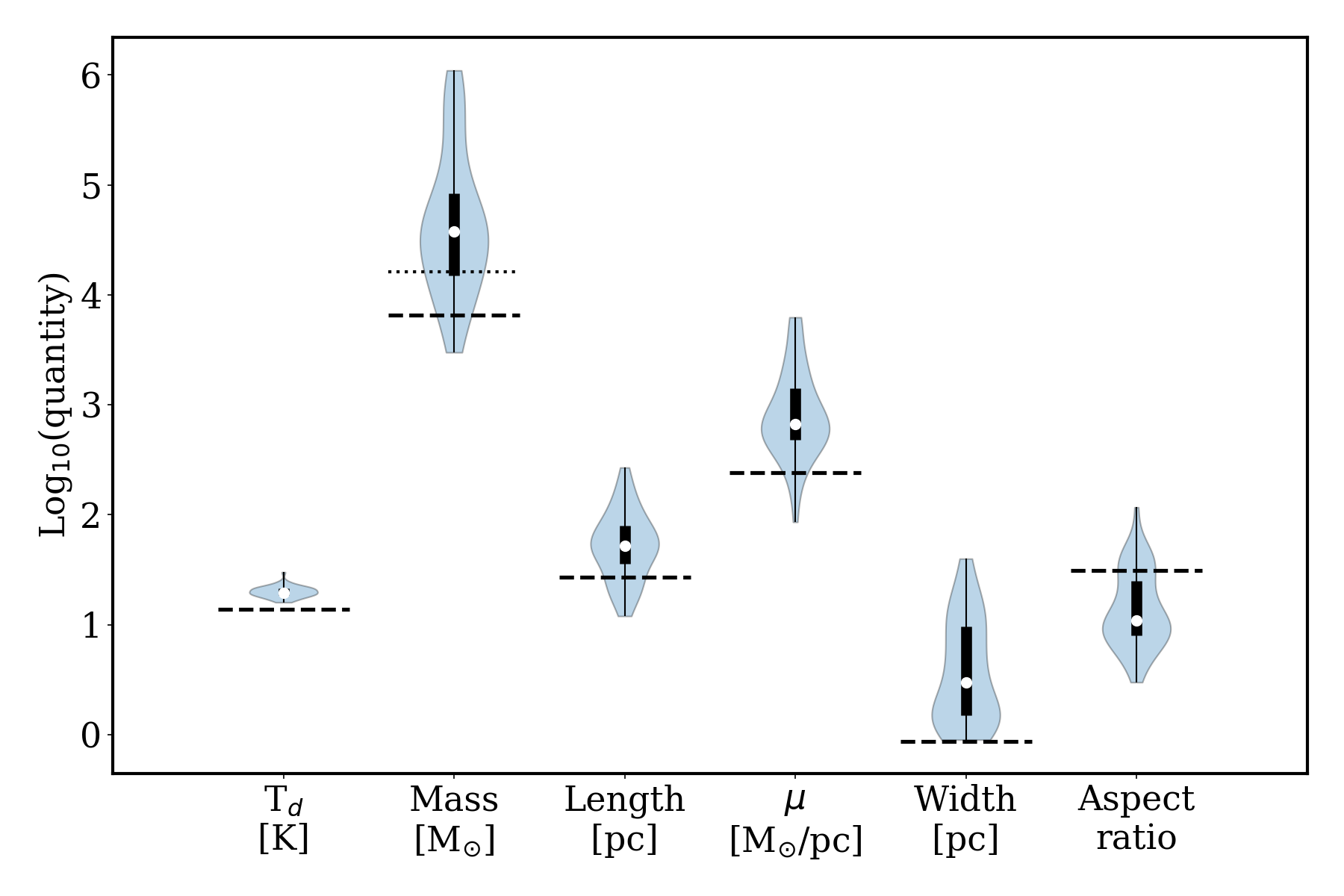}
\caption{A comparison between the properties of G214.5 and those of the GMFs included in the Z18 catalogue. The median property value is shown as a white circle and a black bar shows the interquartile range. The dashed, black, horizontal line in each violin plot shows the property value for G214.5. In the violin plot corresponding to mass, the dashed line is equal to 6570 M$_{\odot}$, the mass within 4 pc of the filament spine defined in section \ref{SSEC:FIL}, and the dotted line is equal to 16,200 M$_{\odot}$, the total mass in the whole G214.5-1.8 region defined in section \ref{SSEC:Prop}.}
\label{fig::compare}
\end{figure}  

Equilibrium models of filament fragmentation result in the prediction of a characteristic fragmentation length-scale, i.e. scales of 0.1 - 2 pc for quasi-periodic core/clump spacing respectively, at approximately 4 times the filament width \citep[e.g. ][]{InuMiy92}.  However, more dynamic models including accretion and turbulence show that there may not be a clear signature of quasi-periodic spacings for multiple reasons such as two-tier hierarchical fragmentation, turbulent dominated fragmentation and/or the presence of sub-filaments \citep{Cla16,Cla17,Cla20}. 

We perform a statistically robust investigation of the possible presence of a characteristic fragmentation length-scale following the recommendations made by \citet{Cla19}, using the nearest neighbour separations (NN), minimum spanning tree separations (MST) and two-point correlation function methods. We use the functions provided in \textsc{FragMent} for this purpose. 

We map the 20 clumps located in the main filament section (C0-C19, seen in figure \ref{fig::spine}) onto the same orthogonal radius-longitudinal distance space as the straightened filament shown in figure \ref{fig::strfil}. We consider only clumps located on, or close to, the filament spine as we may assume that these are clumps formed from the result of filament fragmentation. This is done by taking clumps which are located within 0.7 pc of the spine. This value is chosen as it is approximately twice the filament width, ensuring all clumps close to the filament spine are included. The next closest clump is $\sim$ 1.3 pc away from the spine. Using a 0.7 pc cut, 11 clumps are found with a maximum radial distance of 0.36 pc (locations seen in the bottom panel of figure \ref{fig::strfil} as vertical. dashed lines). For the null hypothesis tests we use a boundary box with a length the same as the filament, $l=27.2$ pc, and a maximum radius of r = $\pm 0.7$ pc. We also use 10,000 runs to sample the null hypothesis distribution that 11 clumps are randomly placed within this boundary box. A minimum separation distance of 0.84 pc is used when generating the null hypothesis distribution as it corresponds to approximately four 18.2'' beams. Such a multiple of the beam size is used to account for the radial extent of the clumps which on average have size of $\sim$ 0.4 pc (see section \ref{SSEC:CLUMPS}). This minimum separation is sufficient as the smallest separation between any two clumps is 1.21 pc.

The results of the three methods may be seen in figure \ref{fig::charlength}. The median and interquartile range of the NN distribution are 1.65 pc and 1.08 pc, and those of the MST distribution are 2.61 pc and 1.39 pc. Both separation distributions are peaked, with the nearest neighbour separation distribution containing a possible second maximum at $\sim3$ pc. One may thus conclude that such a peaked distribution may be indicative of a characteristic fragmentation length-scale but results from the AD test comparing these distributions from those of randomly placed clumps results in $p-$values of 0.087 and $>$0.25 for the NN and MST distributions respectively. Therefore, we cannot detect a statistically significant departure from randomly placed clumps. This is corroborated by the results of the two-point correlation function which is consistently flat and close to zero. There is a possible deficiency of closely spaced clumps below 2 pc but this is likely due to the non-zero size of the clumps. Thus, we find no clear evidence of a characteristic fragmentation length-scale in G214.5 which is consistent with observations of local molecular filament fragmentation in the Gould belt \citep[e.g.][]{And14, Kon15}. It would be of interest to investigate the further fragmentation of the clumps into cores \citep[as has been done in previous works using interferometers, e.g.][]{San19,Tang19,Sah22} as \citet{Kai17} suggest that, with such data, one may discover a two-tier fragmentation pattern. As \citet{Cla19} show that such a pattern may only be robustly detected with a large enough number of cores ($N>20$), the high number of clumps in G214.5 makes it a promising target for such a follow up study.

\subsubsection{Comparing G214.5 to other GMFs}\label{SSSEC:COMPARE}

It is pertinent to compare G214.5 to other known GMFs. As such we compare to the sample of \citet{Zuc18} which comprises 45 GMFs found in the inner Galaxy; hereafter known as the Z18 catalogue. In figure \ref{fig::compare} we compare G214.5's median dust temperature, mass, length, line-mass, width and aspect ratio to the Z18 sample. One can see that G214.5 is comparable to the Z18 filaments in most regards; however, there are two properties which make G214.5 unique. As previously noted in section \ref{SSEC:Prop}, G214.5 is considerably colder than the GMFs found in the Z18 catalogue, likely due to the lower interstellar radiation field in the outer galaxy and the low level of internal heating from the scant number of protostars. Moreover, G214.5 is the narrowest GMF known when compared to the Z18 filaments seen in emission, with a width of 0.87 pc compared to Nessie's 0.90 pc. We note that G214.5 is closer than most of the GMFs studied in Z18 (distance range of 1.9-9.9 kpc), though is at a distance comparable to Nessie, 2.8 kpc. 

\citet{Zha19} present a catalogue of 57 inner galaxy GMFs, with some GMFs also included in the Z18 catalogue, mostly using $^{13}$CO data. Comparing to this dataset we see a similar picture to that seen when comparing to Z18; G214.5 is comparable in its mass and length, though less massive and shorter than the median, but it has an extremely high aspect ratio, $\sim$31 vs. 1.1-6.7.

Comparing to the recent work by \citet{Col21} which covers 37 GMFs in the outer Galaxy, we find that G214.5 has a length consistent with their sample, but due to its narrowness has an aspect ratio greater than any GMF they considered. One finding of \citet{Col21} is that the average mass and line-mass of the GMFs in the outer Galaxy is an order of magnitude below that of those in the inner Galaxy. G214.5 is an exception here and is quite massive in comparison, being more massive than all but 3 GMFs in the \citet{Col21} sample. This exceptionalism is also evident in the fact that G214.5's position is consistent with it lying within the Perseus arm while \citet{Col21} find that nearly all of their sample are associated with inter-arm regions. Thus, G214.5 appears distinct, neither fully resembling an inner or an outer Galaxy GMF. 

\citet{Zuc18} additionally report the radial profile exponent, $p$, from their Plummer-fits of the Bone and large-scale Herschel filament sub-samples and find that they range between 2 and 5.5. Thus they lie separate from the small-scale local filaments studied by \citet{Arz11} which find radial profile exponents in the range 1 to 2.5. In contrast, G214.5 possesses a radial profile exponent of 1.3 - 1.6 (dependent on which side of the radial profile is considered) making it unique amongst GMFs while being typical of smaller-scale filaments. 

Defining the cold and high column density fraction (CCDF) as the number of pixels with a column density above $10^{22}$ cm$^{-2}$ divided by the number with a column density above $3 \times 10^{21}$ cm$^{-2}$ while having a dust temperature below 20 K, \citet{Zuc18} use this quantity in conjunction with the aspect ratio to separate their catalogue sample into 3 populations. These are: elongated giant molecular clouds that appear to be high aspect ratio giant molecular clouds (aspect ratio $\sim$ 8, CCDF $<$ 0.1); elongated dense core complexes which are networks of dense and compact clumps embedded in giant molecular clouds (aspect ratio $\sim$ 10, 0.1 $<$ CCDF $<$ 0.75); and Bone candidates which are highly elongated and dense filaments (aspect ratio $>$ 20, 0.1 $<$ CCDF $<$ 0.5). With an aspect ratio of 31.2, G214.5 is consistent with the Bone candidate population but due to its deficiency of high column density gas it has a CCDF of 0.03, meaning that G214.5 lies outside of these 3 populations. It is unclear what causes the cold and high column density of G214.5 to be so low and for it to lie so far from the Bone candidate population. Connecting this to the low rate of star formation in the cloud, it could be that G214.5 is a young cloud which has not evolved sufficiently to acquire a large amount of dense gas. Or, combined with the asymmetric column density profiles and the evidence for compression, the wider environment, and possible formation mechanism of G214.5, prevents the assembly of a large quantity of dense gas; this is further discussed in the following section.

\section{A HI superbubble compressing G214.5?}\label{SEC:BUB}%

\begin{figure}
\centering
\includegraphics[width=0.95\linewidth]{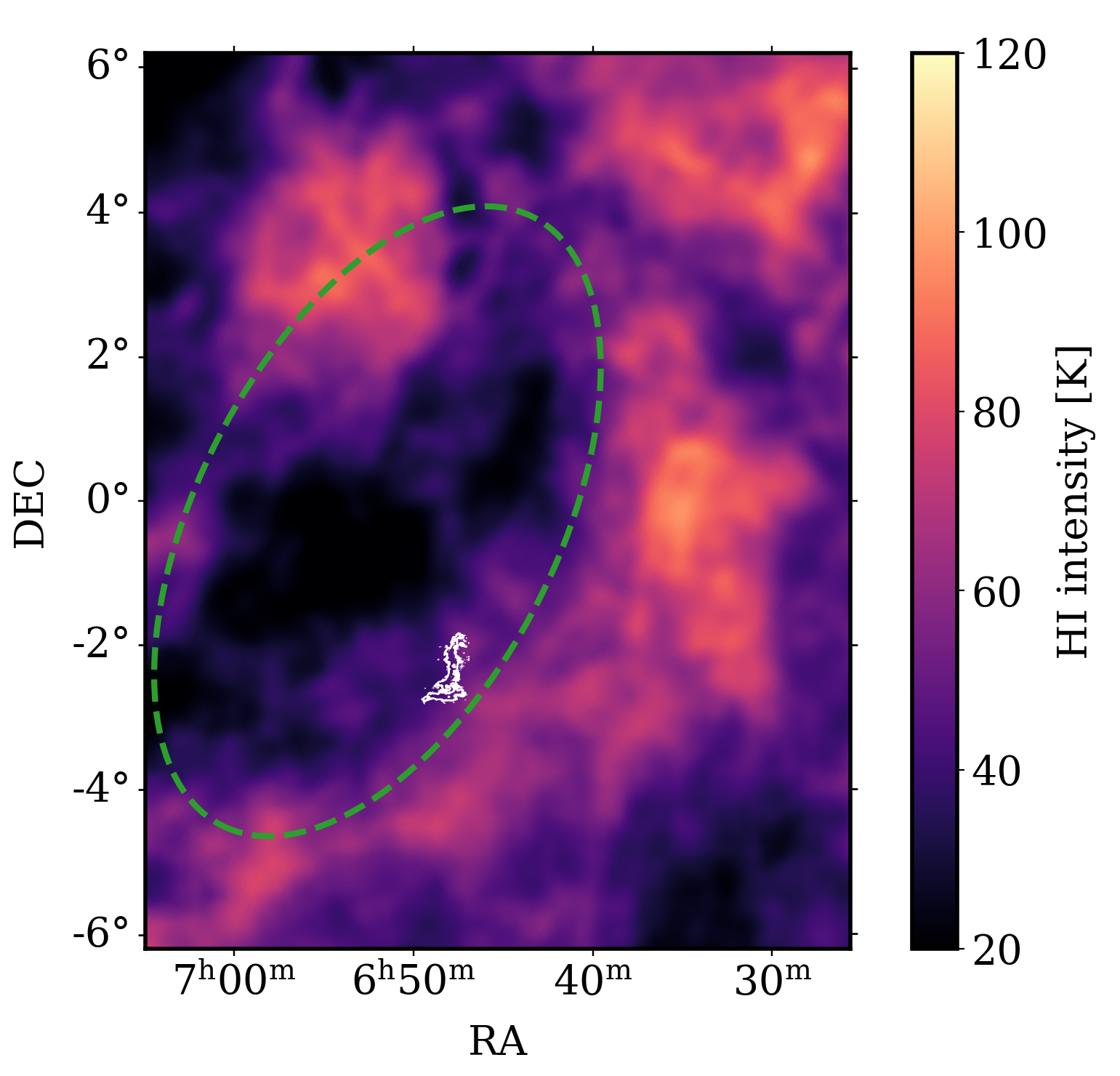}
\includegraphics[width=0.98\linewidth]{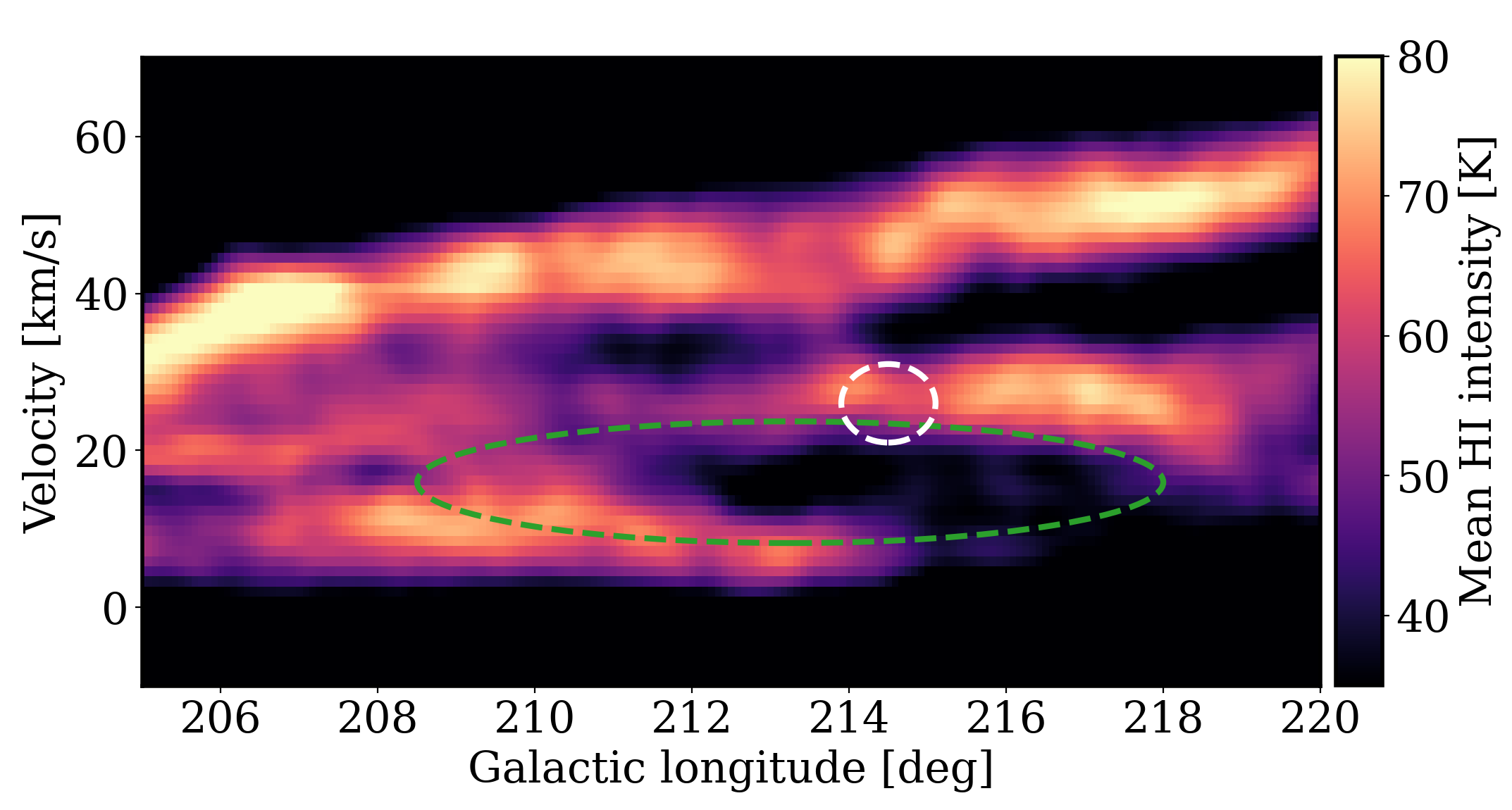}
\caption{(Top) A channel map of the HI 21 cm line intensity at 17 km/s taken showing the environment surrounding G214.5-1.8. The dashed ellipse outlines the HI superbubble GSH214.0+00.0+017.5 identified by \citet{Ehl13} to be centred at 17.5 km/s. The white contour shows the $A_v$ = 1 mag level of the G214.5 GMF. The HI data is taken from the full-sky HI survey HI4PI \citep{HI4PI}. (Bottom) The mean HI brightness in the region $b=\pm3 \degree$ as a position-velocity diagram. The green dashed ellipse shows the HI superbubble's location in position-velocity space and the white dashed ellipse shows the location of G214.5 using the kinematic information from \citet{Yan19}.}
\label{fig::environ}
\end{figure}  

It has long been known that there exist large-scale bubbles within the ISM \citep{Hei79,Hei84,McC02,Chu06,Ehl13}. In HI surveys these features appear as superbubbles with radii above 100 pc which are thought to form due to stellar wind and supernova feedback \citep{McC79,Bru80,McC87,Inu15}. Using the catalogue of HI shells in the Leiden/Argentina/Bonn (LAB) HI survey by \citet{Ehl13} we find that G214.5-1.8 is co-incident, both spatially and kinematically, with the superbubble GSH214.0+00.0+017.5 (figure \ref{fig::environ}). 

\citet{Ehl13} find that this superbubble is located approximately 2 kpc away with a major axis of $\sim$ 350 pc and minor axis of $\sim$ 170 pc, and lies in the velocity range 8.2 - 23.7 km/s, centred at 17.5 km/s. It is thus possible that G214.5 may lie in the far-side of the shell of this superbubble as \citet{Yan19} find that G214.5 lies at a distance of 2.15-2.45 kpc and with a velocity range of $\sim$ 22-32 km/s. The presence of a bubble is apparent in the HI channel map at 17 km/s taken from the full-sky survey HI4PI, shown in the top panel of figure \ref{fig::environ}, which has better angular resolution than the LAB HI survey data used by \citet{Ehl13} to identify it. The kinematic connection can be seen in the HI position-velocity diagram in the bottom panel of figure \ref{fig::environ}, which uses the velocity information from \citet{Yan19} to show the location of G214.5. Furthermore, the bubble is also evidenced by a highly porous morphology seen within the bubble ellipse when observing dust continuum emission from WISE 12 $\mu$m to the Herschel 500 $\mu$m. Additionally, recent work using Gaia and 2MASS data to infer 3D maps of the the Galactic dust distribution find a cavity devoid of significant dust extinction between l $\sim 205\degree$ and l $\sim 220 \degree$ in front of G214.5's location, in agreement with the placement of the superbubble by \citet{Ehl13} \citep[see figure 9 of][]{lal19}. This strong evidence of association means that discussing the consequences of an interaction between the superbubble and G214.5 may be highly fruitful.

The signs of large-scale radial compression in the G214.5 filament shown in section \ref{SSSEC:RADPRO} is highly compatible with the presence of a co-incident, interacting superbubble. In such a hypothetical scenario the large-scale flow driven by the superbubble compresses the main filament structure, most clearly seen in the lower section of the filament, while the head structure has been strongly ablated and torn apart. This would also explain the unique morphology of G214.5: the north-south filament has been compressed from the east and lies completely perpendicular to the head structure which has been disrupted and elongated along the east-west axis by the flow. Moreover, if the filament is mostly compressed by the superbubble while the head structure has predominately been ablated then it could explain the extreme disparity in the location of dense gas in G214.5, 99$\%$ being in the main filament, and thus why star formation is concentrated in the main filament. The origin of why the interaction between G214.5 and the superbubble is different between the filament and the head could be due to a combination of the angle of interaction and a pre-existing north-south density gradient across G214.5; from \citet{Yan19} one sees that the head structure is kinematically closer to the bubble than the main filament, lying at $\sim 22-24$ km/s compared to $\sim 28-32$ km/s.

It is important to also note that the main filament is not purely compressed. As can be seen in the column density plots, there is a large volume of flocculent gas seen to the west of the main filament, predominately in the lower section. \citet{Gol20} show that a shock interacting with a filament may compress a portion of the gas while leading to a steady turbulent stripping of material producing a flocculent wake. This two-sided interaction, in addition to G214.5's potential youth (evidenced by the low star formation activity and low clump luminosity-to-mass ratios), may explain the relatively poor amount of dense gas in G214.5's main filament and why the cloud lies so far from other GMFs in this respect when it is similar in other global quantities, i.e. total mass, length, aspect ratio. Further simulation work investigating such two-sided interaction as well self-gravity and pre-existing density gradients in filaments would be beneficial.

To determine the feasibility of this possible interaction one may consider the timescales of the processes involved. Works studying the interaction of shocks with clouds \citep{Kle94,Xu95} show that the most relative dynamical timescale is the cloud-crushing time, $t_{cc} = r_c \chi^{0.5} / v_s$, where $r_c$ is the equivalent spherical cloud radius, $\chi$ the ratio of cloud density by ambient medium, and $v_s$ is the shock speed in the ambient medium. Taking the area above an $A_v$ of 1 for G214.5 from table \ref{tab::prop} results in an equivalent circle radius of 8.7 pc. To estimate $\chi$ we use the radial profiles determined in section \ref{SSSEC:RADPRO} and take the ratio of the central column density and column density at $r=4$ pc, resulting in a range of approximately $4-10$. \citet{Ehl13} give an expansion speed of $\sim$8 km/s for GSH214.0+00.0+017.5 and we use this for $v_s$. Combined this gives an estimate of the the cloud-crushing time of 2-3 Myr. 

To compare to the cloud-crushing time we calculate the free-fall time for a filamentary cloud using the equation given by \citet{Cla15}: $t_{ff} = (0.49 + 0.26 A_o)/\sqrt{G \rho}$, where $A_o$ is the initial aspect ratio of the filament and $\rho$ the average volume density. The initial aspect is unknown but we may instead take the current calculated aspect ratio of 31.2 to produce a lower limit. To estimate the average density we take the average column density on the spine and divide by the filament width (both taken from table \ref{tab::fil}), resulting in $\rho=6.7 \times 10^{-21}$ g/cm$^{-3}$ ($n\sim 1500$ cm$^{-3}$). This gives a lower limit on the free-fall time of $\sim$ 13 Myr. From this one sees that the cloud-crushing time is sufficiently small compared to the free-fall time that an interaction with the superbubble would dominate the evolution of G214.5 and is fast enough for stripping to occur before gravity may regather material. Moreover, as clouds are fully disrupted by a shock on the order of one to two cloud-crushing times (here a few Myr), G214.5 would have to be relatively youthful for it to be observed in the contiguous state that it is if impacted by the HI superbubble.

To more fully support our hypothesis that G214.5 is impacted by the superbubble GSH214.0+00.0+017.5, kinematic studies are necessary and is the aim of a future paper (Clarke et al. in prep.). Speculating on what one may see if there is an interaction between G214.5 and the superbubble, the resulting compression may appear as a radial velocity gradient across the filament spine while the stripped material may show signs such as higher levels of turbulence. The head structure, if strongly ablated, would consist of lower volume density gas and show a highly turbulent structure. Moreover, careful examination of HI data and CO data in PPV space may further strengthen the association between the superbubble and G214.5, as for example been done by \citet{Daw08} when studying the Carina Flare superbubble's interaction with molecular clouds.  

\section{Conclusions}\label{SEC:CON}%
 
In this paper we have used Herschel data to construct column density and dust temperature maps of the giant molecular filament G214.5-1.8, as well as a 70 $\mu$m source catalogue and a clump catalogue. These data have shown that G214.5 is a unique GMF in multiple ways. 

We show that G214.5 is a giant molecular filament consisting of two distinct regions, a narrow north-south main filament and a perpendicular, flocculent, east-west head structure. Investigating the global properties of G214.5, we find that it is a massive GMF, $\sim16,000$ M$_{\odot}$, yet hosts only 15 potential protostellar 70 $\mu$m sources, making it distinctly quiescent compared to equally massive clouds such as Mon R2 and Serpens. This unusually quiescence, and G214.5's paucity of dense gas, may be due it being a young cloud which has yet to begin star formation in earnest but a more in-depth audit of its star formation history is needed to determine this with certainty.

When investigating the fragmentation of the cloud into clumps, we find a total of 33 clumps with masses ranging from 7 to 227 M$_{\odot}$. These clumps are similar in their general properties to those found in the Hi-GAL sample from \citet{Elia17,Elia21} but fall below the high-mass star formation criterion of \citet{Bal17}, compatible with previous surveys failing to detect any tracers of high-mass star formation in G214.5. Furthermore, we find that 8 of the 33 clumps are associated with a 70 $\mu$m source and are thus protostellar; however, these protostellar clumps are not distributed evenly across G214.5 but predominately located in the main filament (a star-forming fraction of 35$\%$ in the main filament versus 7.6$\%$ in the head structure). We find no statistically significant difference between the clumps in the main filament and the head structure to explain this striking difference in star forming potential, except that the clumps located in the main filament are colder and reach higher peak column densities. From this we infer that the main filament clumps are more deeply embedded and contain a higher proportion of dense gas, both properties promoting star formation. This is supported by the fact that the main filament hosts 99$\%$ of the dense gas ($A_v >$ 7 mag) in G214.5.    

Focusing on the main filament structure, we find that it is approximately 27 pc long and has a width of 0.87 pc, resulting in an aspect ratio of 31. Further, we see that it has a line-mass of 242 M$_{\odot}$/pc, making it highly super-critical and requiring turbulent or magnetic support to remain long-lived. There exist 11 clumps which lie on the filament spine which when studied reveal no statistically significant characteristic fragmentation separation, in agreement with recent filament fragmentation models \citep[e.g.][]{Cla20} and Herschel observations of local filaments \citep[e.g.][]{And14, Kon15}. 

Studying the radial profile of the main filament, we find that the filament may be split into two roughly equal halves with distinct properties, the upper and the lower section. The upper section shows a continuos, cold and dense spine with a narrow and symmetric radial profile (0.71 pc), and the lower section is less dense, warmer, more flocculent and possesses a wide and asymmetric radial profile (2.16 pc). The asymmetry of the lower section is such that the left-hand side of the filament has twice the mass of the right-hand side, and a width over twice the size. The asymmetric radial profiles are highly reminiscent of the profiles of compressed smaller-scale filaments and have comparable asymmetry parameters to those wind-compressed filaments found in the Pipe nebula \citep{Per12}. 

Comparing G214.5 to other giant molecular filaments studied in the \citet{Zuc18} catalogue, we find that it is similar in most global properties to those clouds classified as 'Bone' candidates (highly elongated and massive filaments) but is both colder and narrower than any other. While being most similar to these 'Bone' candidate GMFs out of the 3 population groups suggested by \citet{Zuc18}, G214.5 differs significantly from them due to its low cold and high column density fraction: 0.03 compared to the typical range of 0.1 - 0.5. Comparing to the outer Galaxy GMFs studied by \citet{Col21}, we find that G214.5 is comparable in size but is one of the most elongated and massive GMFs.

Considering the wider environment of G214.5, we find evidence that the GMF lies at the edge of the HI superbubble GSH214.0+00.0+017.5 found by \citet{Ehl13}. From considering the distances to the bubble and G214.5, as well as the kinematics, we suggest that G214.5 lies on the far-side of the superbubble's shell. A hypothetical interaction of G214.5 with a superbubble has the potential to explain the large-scale compression seen as well as the unique morphology of a dense, compressed main filament lying perpendicular to an eroded, diffuse head structure. From considering the relevant timescales we show that it is possible for such an evolution to occur within a fraction of a cloud free-fall time. It is unclear whether G214.5 may have formed in the shell of the superbubble, or if it was a pre-existing structure which has been impacted by the shell's expansion. Both scenarios are consistent with the bubble-driven ISM picture presented by \citet{Inu15} which emphasises the important role of large-scale feedback bubbles in promoting cloud formation. We wish to highlight that while such bubbles may compress material, G214.5 is a potential example where they may simultaneously erode gas from structures leading to a complex interplay of compression and ablation which has consequences on the resulting star formation; a scenario supported by the simulations of \citet{Gol20}. 

We conclude that G214.5-1.8 is a cold, massive and quiescent giant molecular filament which is potentially still young and in an early stage of its evolution. While being similar to the highly elongated and massive 'Bone' candidate filaments, G214.5 is unique  in its paucity of dense gas which is likely connected to its quiescence and possible youth. A potential interaction with a HI superbubble may have lead to the large-scale compression of parts of the GMF while simultaneously ablating other regions, resulting in uneven star formation activity across the cloud. Thus, G214.5 may serve as an example of the intersection of the bubble-driven ISM paradigm of \citet{Inu15} and the filament paradigm of star formation presented by \citet{And14}, an idea developed in the upcoming PPVII chapter \citet{Pin22}, making it a unique and interesting object worthy of further study.

\section{Acknowledgments}\label{SEC:ACK}%
SDC would like to thank Ya-Wen Tang and Ana Duarte-Cabral for discussions which have improved the paper. The authors also thank the anonymous referee for their helpful comments. SDC is supported by the Ministry of Science and Technology (MoST) in Taiwan through grant MoST 108-2112-M-001-004-MY2. ASM acknowledges support from the Collaborative Research Centre (SFB 956, sub-project A6), funded by the Deutsche Forschungsmeneinschaft (DFG). GMW acknowledges support from the UK’s Science and Technology Facilities Council (STFC) under grant number ST/W00125X/1. ADPH gratefully acknowledges the support of a PhD studentship from the UK Science and Technology Facilities Council (STFC).  SW acknowledges support from the ERC starting grant No. 679852 `RADFEEDBACK' and thanks the DFG for funding through the Collaborative Research Center (SFB956) on the `Conditions and impact of star formation' (sub-project C5). NS acknowledges support by the Agence National de Recherche (ANR/France)
and the Deutsche Forschungsgemeinschaft (DFG/Germany) through the project `GENESIS' (ANR-16-CE92-0035-01/DFG1591/2-1).

\section{Data availability}\label{SEC:DATA}%
The data underlying this article will be shared on reasonable request to the corresponding author. The analysis tool of this work, \textsc{FragMent}, is made freely available at https://github.com/SeamusClarke/FragMent.

\bibliographystyle{mn2e}
\bibliography{ref} 

\begin{appendices}

\section{Herschel band images}\label{APP:HBAND}%
\begin{figure*}
\centering
\includegraphics[width=0.48\linewidth]{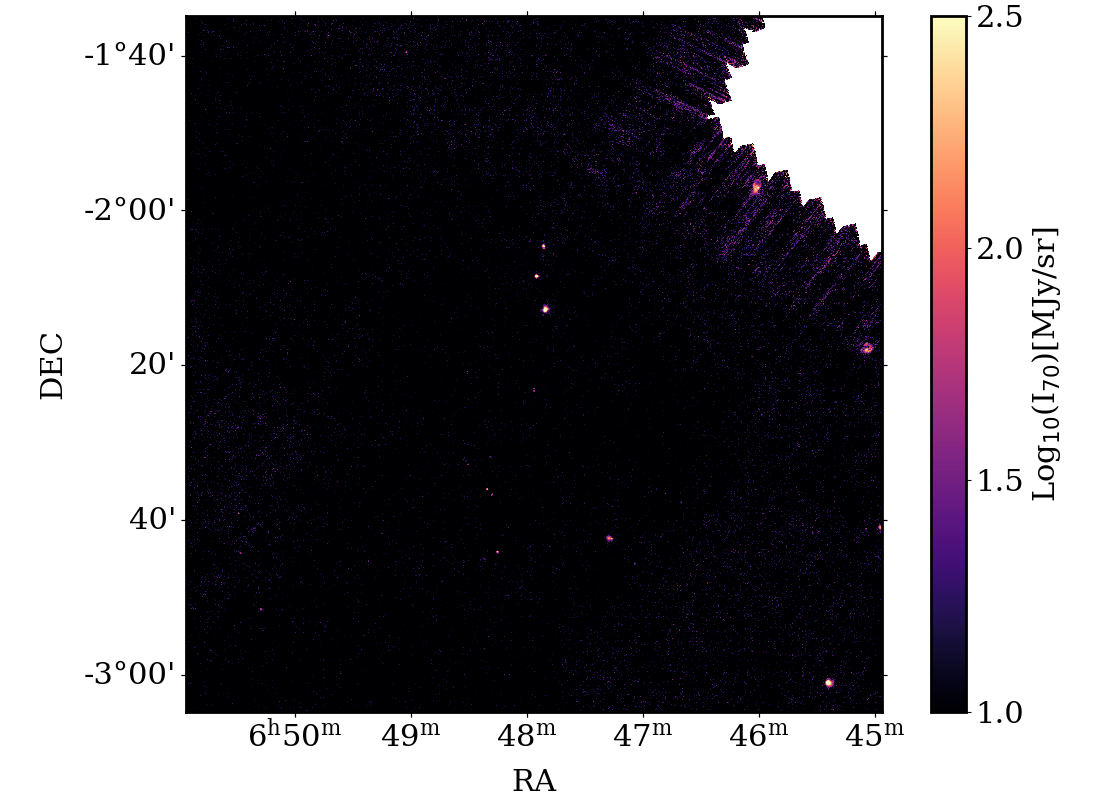}
\includegraphics[width=0.48\linewidth]{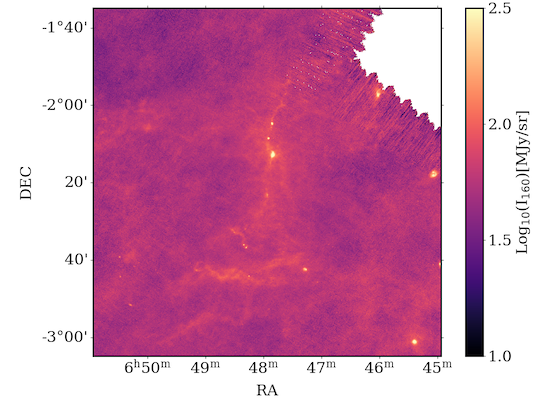}
\includegraphics[width=0.48\linewidth]{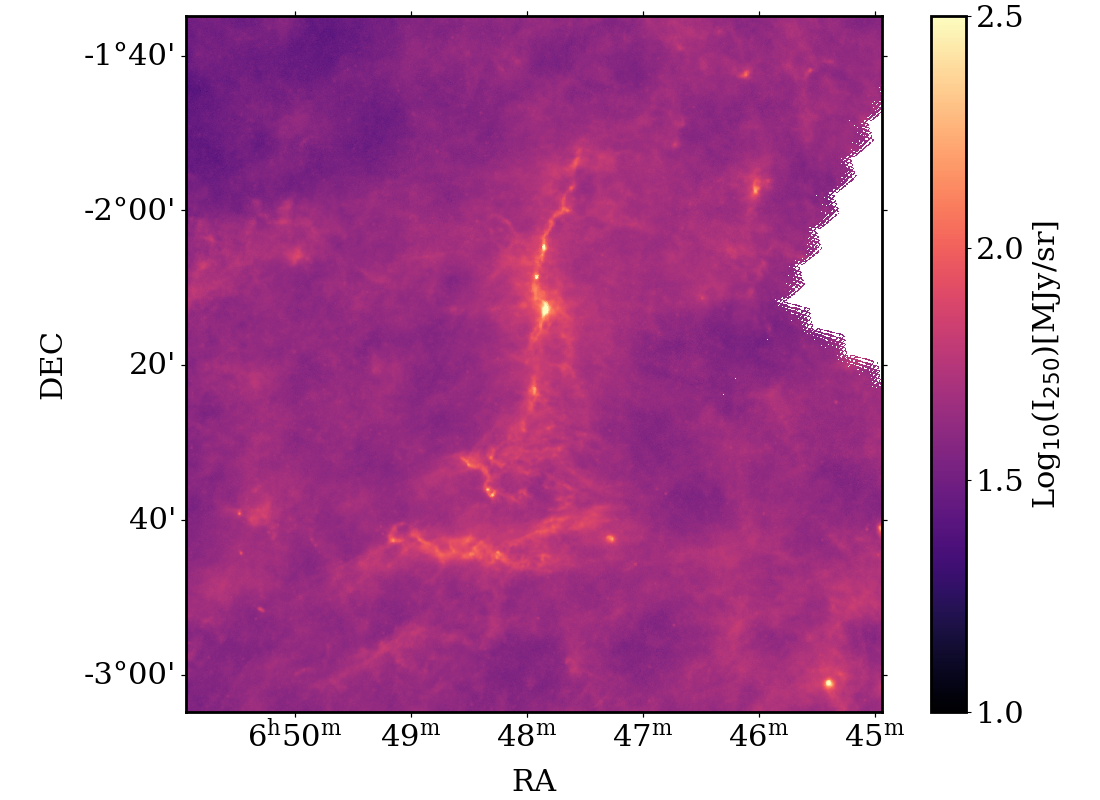}
\includegraphics[width=0.48\linewidth]{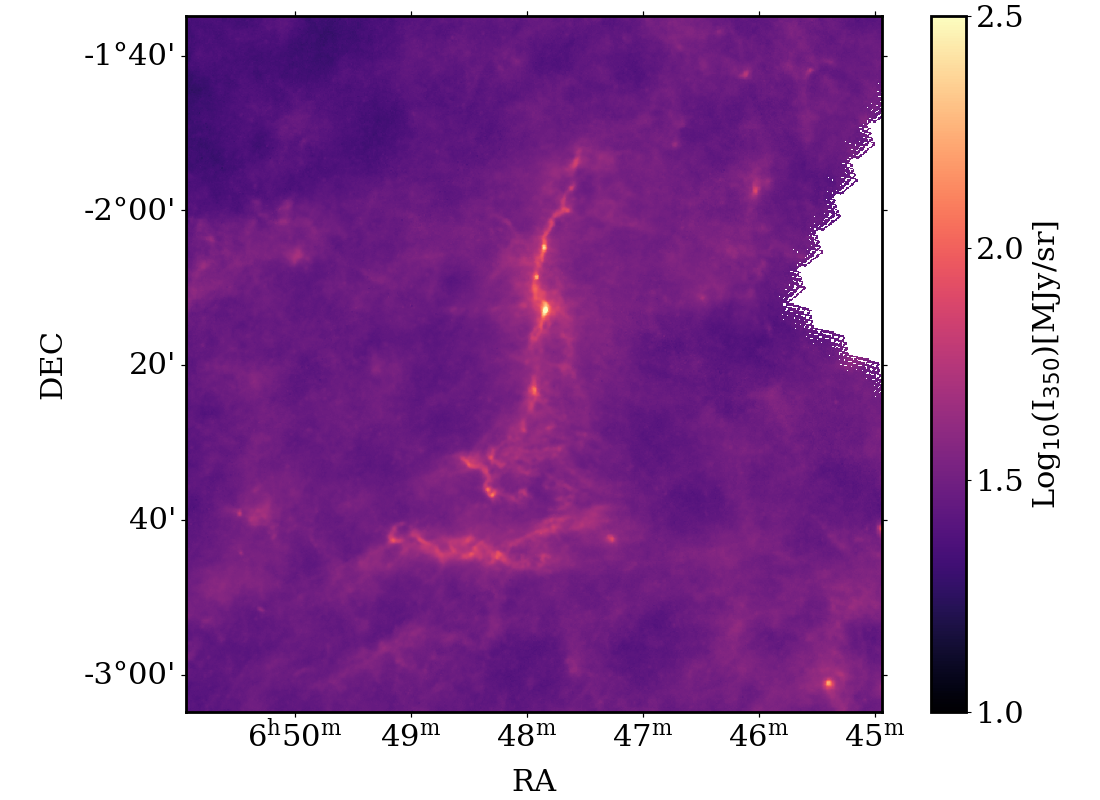}
\includegraphics[width=0.48\linewidth]{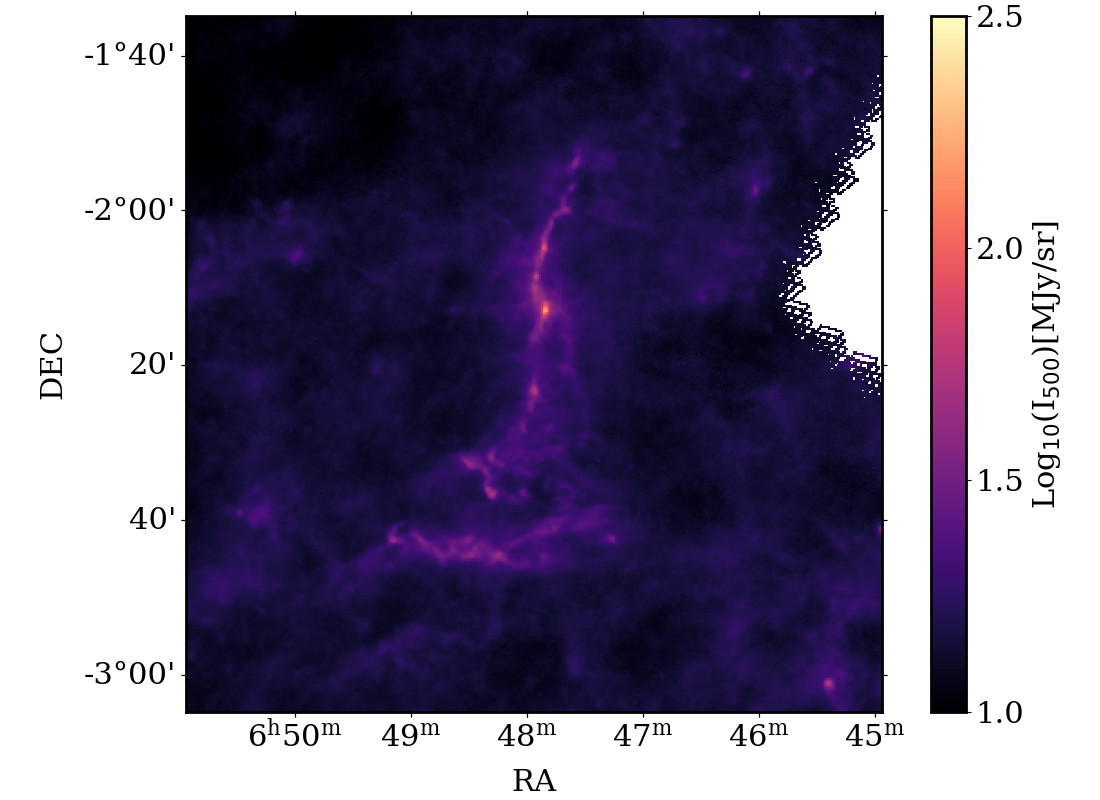}
\caption{Herschel band images of the (top left) 70 $\mu$m, (top right) 160 $\mu$m, (middle left) 250 $\mu$m, (middle right) 350 $\mu$m and (bottom) 500 $\mu$m emission. The intensity is shown in base 10 logarithm to enhance the visibility of the surrounding emission, and all colour bars are set to the same scale for comparison between the bands. }
\label{fig::appband}
\end{figure*}

\begin{figure}
\centering
\includegraphics[width=0.98\linewidth]{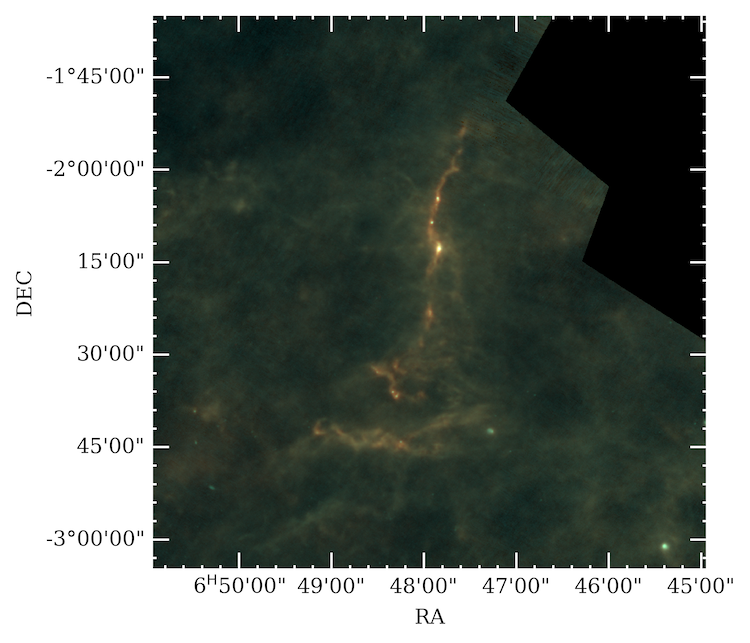}
\caption{ A RGB image (R: 500 $\mu$m, G: 250 $\mu$m, B: 160 $\mu$m) of G214.5. The colour bar of each band is logarithmically scaled and ranges from the minimum to the maximum of that band.}
\label{fig::RGB}
\end{figure}

\numberwithin{table}{section}
In figure \ref{fig::appband} we show the maps of the 70, 160, 250, 350 and 500 $\mu$m Herschel images used to produce the column density and dust temperature maps in section \ref{SSEC:CDTEMP}. Figure \ref{fig::RGB} is an RGB image of G214.5 to help compare the distributions of the Herschel emission in different bands. The image was made using the \textsc{Python} package \textsc{MultiColorFits} \citep[][https://github.com/pjcigan/multicolorfits]{Cig19}.

\newpage

\section{Clump comparison with Elia et al. (2021)}\label{APP:ELIA}%
\begin{figure}
\centering
\includegraphics[width=0.98\linewidth]{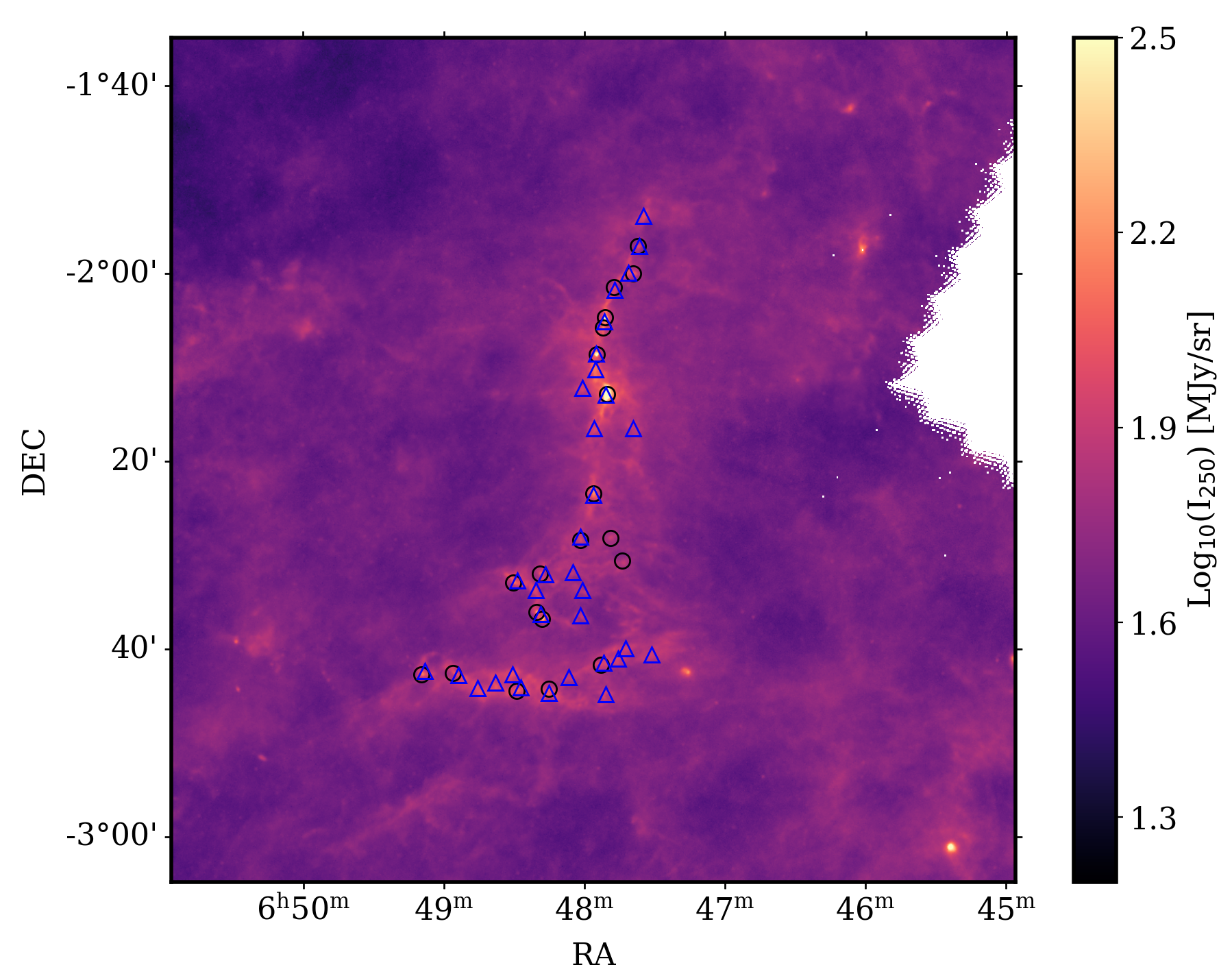}
\caption{A 250 micron image with the centres of the dendrogram clumps identified in this work (section \ref{SSEC:CLUMPS}) overlaid as blue triangles, and the Hi-GAL clumps identified by \citet{Elia21} overlaid as black circles.}
\label{fig::compareElia}
\end{figure}

\numberwithin{table}{section}
In this appendix we compare the properties of the clumps identified in this work (hereafter dendrogram clumps) with the co-incident clumps identified by \citet{Elia21} (hereafter referred to as Hi-GAL clumps). Figure \ref{fig::compareElia} shows the 250 micron image of G214.5 overlaid with the 33 clumps identified in this work (blue triangles) and the 20 Hi-GAL clumps identified (black circles) which lie in this area. In general one sees good agreement between the two samples. Nearly all Hi-GAL clumps are identified as dendrogram clumps in this work, though there are a number of dendrogram clumps which are not associated with Hi-GAL clumps. This difference in distribution is due to the very different identification techniques: a dendrogram constructed from the column density map vs. CuTEx identified peaks in Herschel band images which are subsequently cross-matched. 

\begin{figure}
\centering
\includegraphics[width=0.97\linewidth]{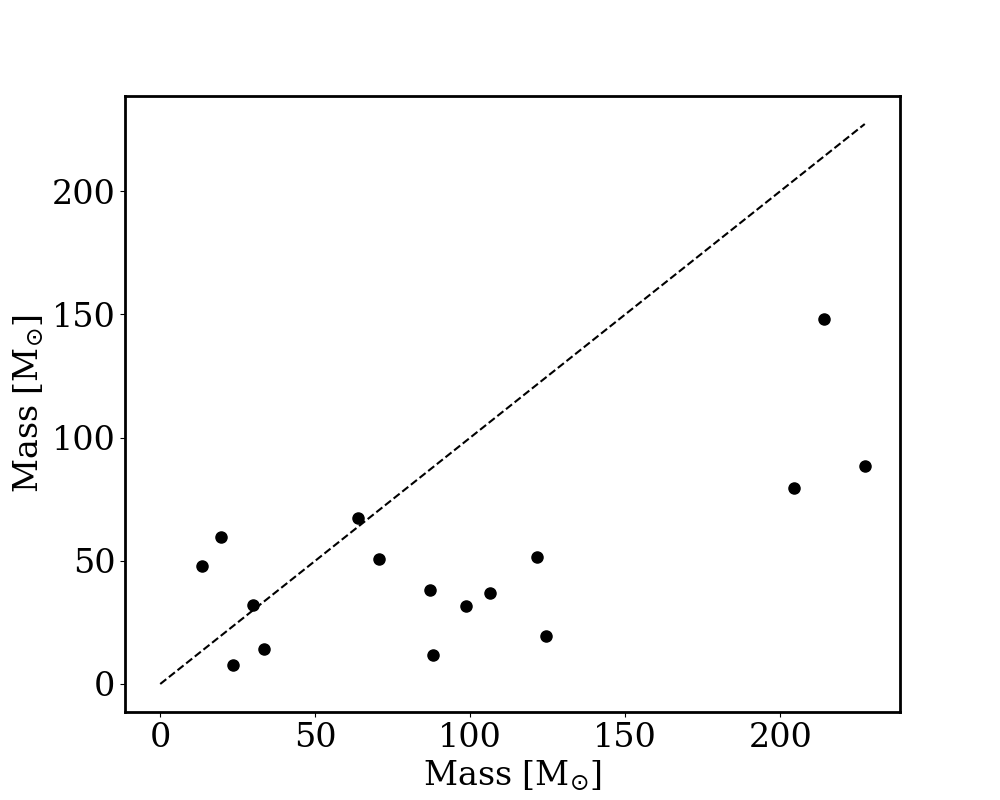}
\includegraphics[width=0.97\linewidth]{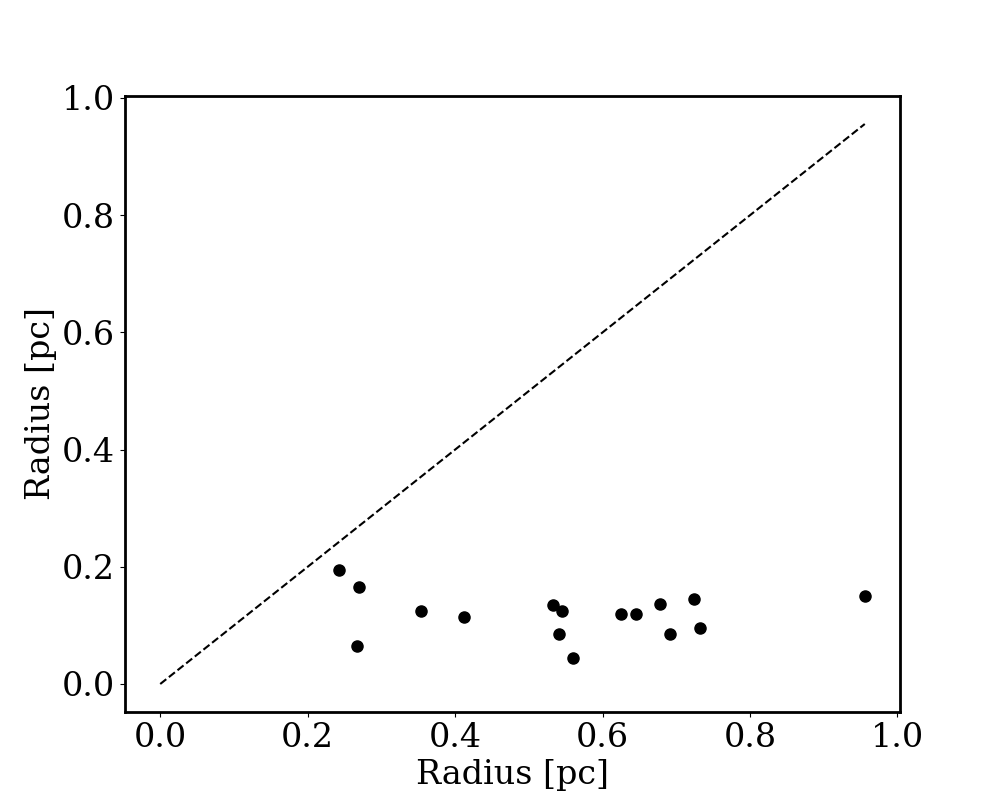}
\caption{(Top) The mass of the 16 co-incident clumps as measured in this work ($x$-axis) and in \citet{Elia21} ($y$-axis). (Bottom) The radius of the 16 co-incident clumps as measured in this work ($x$-axis) and in \citet{Elia21} ($y$-axis). The dashed line in both panels shows the 1:1 ratio.}
\label{fig::mr_compare}
\end{figure}

\begin{figure}
\centering
\includegraphics[width=0.97\linewidth]{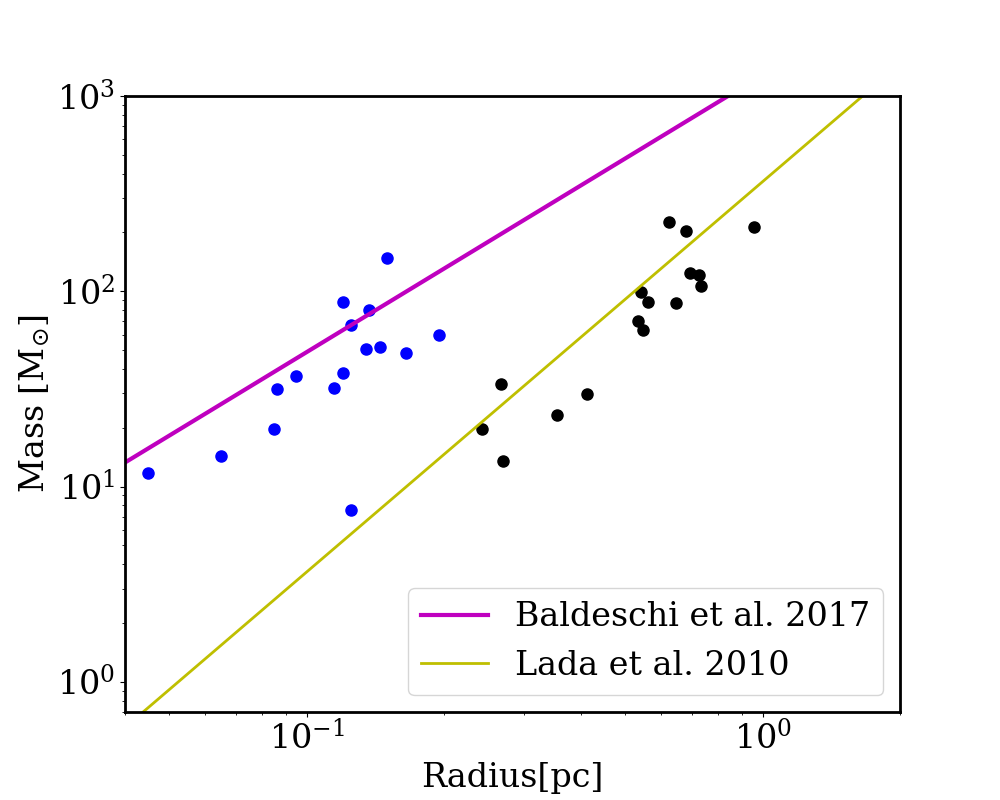}
\includegraphics[width=0.97\linewidth]{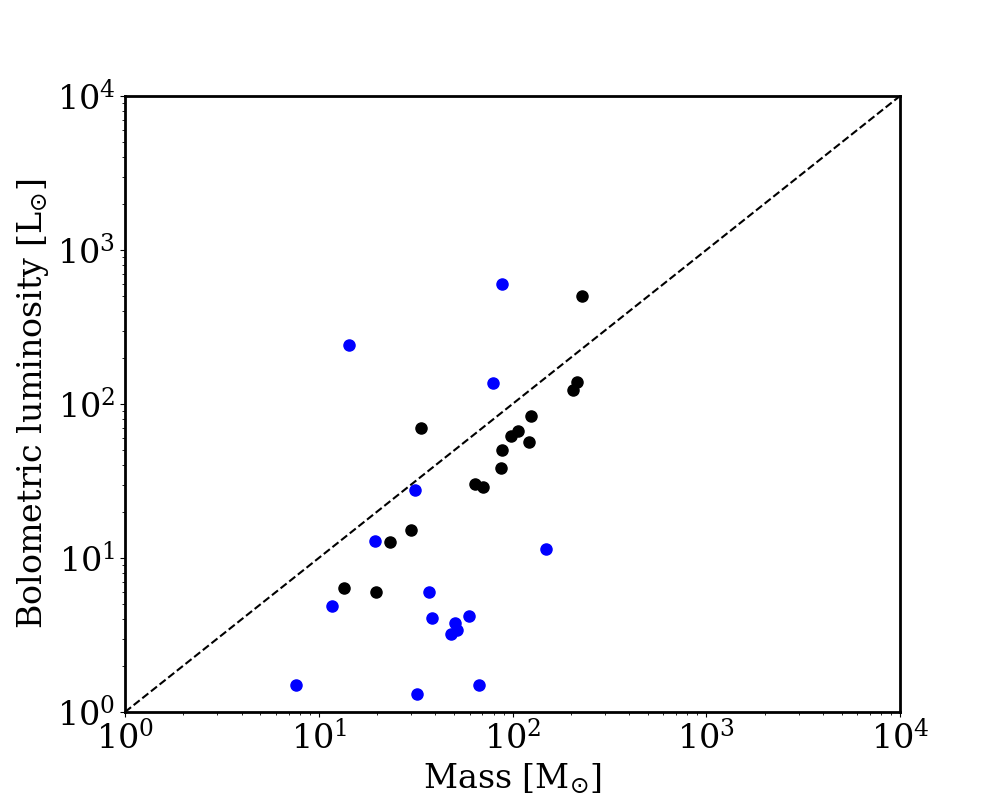}
\caption{(Top) The mass-radius relation for the 16 co-incident clumps as measured in this work (black circles) and as measured by \citet{Elia21} (blue circles). Also shown are the (purple) high-mass star formation threshold from \citet{Bal17} and the (yellow) efficient star formation threshold from \citet{Lada10}. (Bottom) The mass-luminosity relation for the 16 co-incident clumps as measured in this work (black circles) and as measured by \citet{Elia21} (blue circles) with the 1:1 line shown as a dashed line.}
\label{fig::mrlum_compare}
\end{figure}

There are 16 dendrogram clumps with co-incident Hi-GAL clumps, defined as having centres within 1 arcminute of each other. Two of the dendrogram clumps are co-incident with 2 Hi-GAL clumps: for comparison the mass, bolometric luminosity, and size of the two Hi-GAL clumps is summed, while the mean of the two is taken for all other quantities. Figure \ref{fig::mr_compare} shows the comparison between the mass of the co-incident clumps derived in this work and the mass derived in \citet{Elia21}, as well a comparison of the clump radii \footnote{The \citet{Elia21} catalogue does not provide uncertainty estimates for clump radii or bolometric luminosity, only for their mass. For this reason we do not show the uncertainties in figures \ref{fig::mr_compare} and \ref{fig::mrlum_compare}}. One can see that the sizes of the Hi-GAL clumps are considerably smaller than the dendrogram clumps, with sizes close to the beam size ($\sim 0.11$ pc). This is because CuTEx fits two-dimensional Gaussians to identified peaks in the Herschel emission maps, thus only the inner brightest region is fitted which is comparable to the beam size. Further, \citet{Elia21} compute deconvolved radii allowing for sizes below the beam. This is not done in this work because of the elongated nature of a number of the clumps making it non-obvious how best to do this deconvolution. Moreover, because the area over which the clump is defined is much smaller for the Hi-GAL clumps they typically have smaller masses than the dendrogram clumps. 

In figure \ref{fig::mrlum_compare} one sees the effect of the different identification techniques has on the placement of the clumps in the mass-radius plot and the mass-luminosity plot. In the mass-radius plot, the Hi-GAL values place the clumps considerably to the left due to their reduced size, and thus all clumps now lie above the \citet{Lada10} threshold. This shift shows the degree of care which must be taken when comparing thresholds and clumps identified from different studies due to the different techniques employed, a point raised by \citet{Elia21}. In the mass-luminosity plot, most of the Hi-GAL derived bolometric luminosities are lower than those derived here in this work. This is because the majority of the clumps do not have far-infrared counter-parts (i.e. $<70$ micron sources) and thus their bolometric luminosity is strongly linked to their size (as seen in figure \ref{fig::corr}). Regardless of if the Hi-GAL or dendrogram clump values are considered, the conclusion that the majority of the clumps have low luminosity-to-mass ratios, and are thus evolutionarily young, is the same.

\end{appendices}

\label{lastpage}

\end{document}